\documentclass[11pt,a4paper]{article}
\pdfoutput=1

\usepackage[dvipdfmx]{graphicx}
\usepackage{bmpsize}
\usepackage{jheppub}

\usepackage{ascmac}

\usepackage{epsf}
\usepackage{color}
\usepackage{amsmath}
\usepackage{amssymb}
\usepackage{latexsym}
\usepackage{braket}
\usepackage{slashed}
\usepackage{float}
\usepackage{cases}
\usepackage{bm}
\usepackage{arydshln}

\newcommand{\A}{\mathcal{A}}

\newcommand{\D}{\mathcal{D}}

\newcommand{\al}[1]{\begin{align}#1\end{align}}

\newcommand{\bp}{\begin{pmatrix}}
\newcommand{\ep}{\end{pmatrix}}
\newcommand{\bb}{\begin{bmatrix}}
\newcommand{\eb}{\end{bmatrix}}

\newcommand{\del}{\partial}

\usepackage[normalem]{ulem}

\newcommand{\beq}{\begin{equation}}
\newcommand{\eeq}{\end{equation}}
\newcommand{\bea}{\begin{eqnarray}}
\newcommand{\eea}{\end{eqnarray}}

\newcommand{\fulltoday}{\number\day\space \ifcase\month\or
    January\or February\or March\or April\or May\or June\or
    July\or August\or September\or October\or November\or December\fi
    \space\number\year}

 \makeatletter
    
    \@addtoreset{equation}{section}
  \makeatother

\allowdisplaybreaks[3]

\usepackage{calc}
\newcounter{hours}\newcounter{minutes}
\makeatletter                   
\renewcommand*{\thehours}{\two@digits\c@hours}
\renewcommand*{\theminutes}{\two@digits\c@minutes}
\makeatother

\title{\boldmath Active Dirac Neutrinos via $SU(2)_L$ Doublets in 5d}

\author[a]{Yukihiro Fujimoto,}
\author[b]{K. Hasegawa,}
\author[c]{Tomoaki Nagasawa,}
\author[d,1]{Kenji Nishiwaki,\note{Corresponding author.}}
\author[b]{Makoto Sakamoto,}
\author[b]{Kentaro Tatsumi\,}

\affiliation[a]{National Institute of Technology, Oita College, Oita 870-0152, Japan}
\affiliation[b]{Department of Physics, Kobe University, Kobe 657-8501, Japan}
\affiliation[c]{National Institute of Technology, Tomakomai College, Tomakomai 059-1275, Japan}
\affiliation[d]{School of Physics, Korea Institute for Advanced Study, Seoul 02455, Republic of Korea}

\emailAdd{y-fujimoto@oita-ct.ac.jp}
\emailAdd{kouhei@phys.sci.kobe-u.ac.jp}
\emailAdd{nagasawa@tomakomai-ct.ac.jp}
\emailAdd{nishiken@kias.re.kr}
\emailAdd{dragon@kobe-u.ac.jp}
\emailAdd{134s110s@stu.kobe-u.ac.jp}

\abstract{
We propose a new mechanism to generate minuscule active neutrino masses in a five-dimensional~(5d) spacetime of an interval without introducing $SU(2)_L$ singlet neutrinos.
Under asymmetric boundary conditions on the two end points, a bulk mass for a 5d fermion allows a Dirac particle with a tiny mass eigenvalue.
Implementing this mechanism, {which provides us a new tool for building neutrino mass models,} to the standard model gauge structure is possible when all the gauge bosons and the Higgs boson are localized on one of the branes.
}

\subheader{Report number: KIAS-P16007, KOBE-TH-16-01}
\arxivnumber{1601.05265}

\begin{document}

\maketitle
\flushbottom

\section{Introduction}

One of the recent greatest experimental triumphs in particle physics is the confirmation of the nonzero neutrino mixing angle $\theta_{13}$~\cite{An:2012eh,Ahn:2012nd,Adamson:2013whj,Abe:2013hdq,Abe:2014lus}.
After combining these data through suitable statistical methods, the three mixing angles of the neutrino mixing matrix $(U_{\text{PMNS}})$, proposed by Pontecorvo, Maki, Nakagawa, and Sakata~(PMNS)~\cite{Maki:1962mu,Pontecorvo:1967fh},
were precisely pinned down~\cite{Tortola:2012te,Fogli:2012ua,GonzalezGarcia:2012sz,Capozzi:2013csa,Gonzalez-Garcia:2014bfa,Bergstrom:2015rba,Gonzalez-Garcia:2015qrr}.
Also, the experimental results and the results of the global analyses provide us the information on the Dirac phase, which describes the CP violation in the lepton sector, even though all the possibilities of the phase are still consistent within a $3\sigma$ confidence level~\cite{Bergstrom:2015rba,Gonzalez-Garcia:2015qrr}.
When we measure the CP phase and determine the pattern of the ordering in neutrino masses (normal or inverted), we achieve a comprehensive understanding on the nature of neutrinos.

On the other hand, we should provide a reasonable answer to the question, ``why (active) neutrino masses are {so} tiny?''.
These mass spectrum should be almost degenerated and the sum of the eigenvalues are {constrained} by Planck experiment~\cite{Ade:2013zuv}, concretely speaking it being less than $0.23\,\text{eV}$.
The simplest extension of the standard model~(SM) for realizing the experimental result is to introduce new $SU(2)_L$ singlet right-handed neutrinos and Dirac mass terms for the neutrino masses.
Unfortunately in this {extension}, the tiny mass eigenvalues should be realized by hand.

Around 1980, the attempt to build the neutrino mass model
which can explain the smallness in a natural way, began.
In the grand unified theory (GUT), the lepton number 
is violated in general and {Majorana neutrino masses can exist}.
Inspired by the low energy effective theory of the GUTs,
Majorana neutrino mass models were built
as minimal extensions of the SM.
Three pioneer works, seesaw model~\cite{Minkowski:1977sc,Yanagida:see-saw,Gell-Mann:see-saw,Mohapatra:1979ia}, 
$SU(2)_L$ triplet Higgs model~\cite{Cheng:1980qt,Schechter:1980gr,Schechter:1981cv},
and Zee model~\cite{Zee:1980ai}, 
were submitted.
In the three models, the lepton number is explicitly broken
due to newly introduced fields and interactions,
and Majorana mass terms {for the left-handed $SU(2)_{L}$ doublet neutrinos are induced}.
The smallness of the neutrino masses is naturally explained
by the seesaw mechanism {in} seesaw and $SU(2)_L$ triplet Higgs models,
and by the one loop amplitude {in} Zee model.
After that, many progresses and variations were made,
which are reviewed in, for instance, 
\cite{Mohapatra:1998rq,Fukugita:2003en,Ma:2009wt,Altarelli:2014dca}.
We note that the building of loop-induced neutrino scenarios is recently active~\cite{Zee:1985id,Babu:1988ki,Krauss:2002px,
Ma:2006km,Aoki:2008av,Gustafsson:2012vj,Hatanaka:2014tba,Nishiwaki:2015iqa,Chao:2015nac,Kanemura:2015bli,Nomura:2016fzs,Yu:2016lof,Ding:2016ldt,Nomura:2016seu,Okada:2016rav}.
Experiments have not yet determined whether the neutrino masses obey the Dirac or Majorana type.
Neutrino oscillation experiments cannot determine the type.
Neutrinoless double beta decay is one signal of
Majorana neutrino.
The Heidelberg–-Moscow experiments reported the
signal~\cite{KlapdorKleingrothaus:2000sn,KlapdorKleingrothaus:2001ke},
but the results have not yet been confirmed~\cite{Agashe:2014kda}.
The possibility of pure Dirac type is not excluded.
Then in the present article, we pursuit a neutrino mass model
of pure Dirac type.

Crossing the last millennium, new solutions to the gauge hierarchy problem
were {suggested}, which are constructed in higher dimensional spacetime including
brane structure.
They are {large extra dimensions}~\cite{ArkaniHamed:1998rs,Antoniadis:1998ig} and 
Randall-Sundrum~(RS) models~\cite{Gogberashvili:1998vx,Randall:1999ee}.
Based on the frameworks, many models to explain
the fermion mass hierarchy 
appeared~\cite{ArkaniHamed:1998vp,ArkaniHamed:1999dc,Dvali:1999cn,Yoshioka:1999ds,Mohapatra:1999zd,
Grossman:1999ra,Gherghetta:2000qt,Huber:2000ie,
Moreau:2004qe,Moreau:2005kz,Frank:2014aca,Frank:2015sua}.
A successful model where tiny and pure Dirac mass terms are generated in a natural way is the model by
Grossman-Neubert~\cite{Grossman:1999ra}.
The model is based on the RS spacetime geometry.
Only the graviton and the $SU(2)_L$ singlet right-handed neutrino
fly in the bulk, and all the SM fields are confined in the TeV brane.
The right-handed neutrino is localized near the Planck brane
and the value of the mode function on the TeV brane
is suppressed by the warp factor.
Then the tiny Dirac mass terms are naturally induced
through the Yukawa couplings on the TeV brane.
Apart from above frameworks, various works had been made for addressing issues
related to flavor structure in the context of extra dimensions~\cite{
Dvali:2000ha,Shaposhnikov:2001nz,Neronov:2001qv,Kaplan:2001ga,
Libanov:2000uf,
Frere:2000dc,
Frere:2001ug,
Frere:2003hn,
Cremades:2004wa,
Nagasawa:2004xk,Parameswaran:2006db,Abada:2006yd,Gogberashvili:2007gg,
Park:2009cs,Kusenko:2010ik,Csaki:2010az,Fujimoto:2011kf,
Kaplan:2011vz,Fujimoto:2012wv,Fujimoto:2013ki,
Fujimoto:2013xha,Takahashi:2013eva,Abe:2013bca,Cai:2015jla,
Fujimoto:2014fka,Abe:2014noa,Abe:2015yva}.
Among them, {a series of models are} constructed on the flat extra dimension
of {an interval~\cite{Fujimoto:2012wv} or $S^{1}$~\cite{Fujimoto:2013ki,Fujimoto:2014fka},} and the compactification scale is taken as traditional small one.
The {models introduce} some point interactions (zero-thickness branes)
and can derive the SM plus the observed neutrino masses and mixings.
All fields live in the bulk, {including}
singlet fermions which become
the $SU(2)_L$ singlet right-handed neutrinos 
in four dimension after the Kaluza-Klein decomposition.
Then, the tiny and Dirac neutrino masses are induced 
without fine tuning.\footnote{
We mention that this direction was also applied for the generation of Majorana mass terms~\cite{Cai:2015jla}.}

{In contrast with the above models,
in this article, we present a new mechanism
to induce pure Dirac neutrino masses on the small and flat
extra dimension without any $SU(2)_L$ singlet right-handed neutrino or other fields for radiative generation of Majorana mass terms.
The mechanism might be the simplest one among
mechanisms to induce tiny pure Dirac masses naturally.
This new mechanism is discussed {again} on five-dimensional (5d) space-time of an interval.}
A key point is that we consider asymmetric boundary conditions {(BC's)} on the two end points.
As we see later, in a certain parameter choice, both of left and right components of the active neutrinos are provided as 4d states of a 5d $SU(2)_L$ doublet neutrino with {a tiny mass}, which is described as the a fundamental mass scale times {an exponential} suppression factor.

We propose also a prototype of a realistic model in which the new mechanism is embedded.
A nontrivial point is that we also predict {active} right-handed {components} under $SU(2)_L$, which seems to lead to additional gauge interactions, and eventually new contributions to the invisible decay width of the $Z$ boson, which are severely restricted by the LEP experiments~\cite{Barate:1999ce,Acciarri:2000ai,Abreu:2000mh,Abbiendi:2000hu,Collaborations:2000aa,ALEPH:2002aa}.
We show that implementing this mechanism to the standard model gauge structure is possible when all the gauge bosons and the Higgs {boson} are localized on one of the branes, where the right-hand components have almost zero overlaps with the $Z$ boson, and thereby we can evade the constraint from the invisible decay channel.

This paper is organized as follows.
In Sec.~\ref{sec:mass_spectrum}, we see details on the configuration which generates {tiny Dirac mass} under asymmetric {BC's}.
In Sec.~\ref{sec:implementation}, we discuss how to implement the above mechanism to the SM gauge structure consistently.
In Sec.~\ref{eq:conclusion}, we summarize our results and conclude.

\section{Boundary Condition for Dirac Neutrino with Minuscule Mass
\label{sec:mass_spectrum}}

In this section, we revisit the setup discussed in Ref.~\cite{Fujimoto:2011kf}, where we investigate a 5d free fermion with a bulk mass $M$.\footnote{
{Note that a similar discussion is found in Ref.~\cite{Nagasawa:2004xk}.}
}
The action is given by
\al{
S_{\text{f,free}} =
\int \!\! d^4x \int_0^L \!\!\! dy \left\{ \overline{\Psi}(x,y) \left[ i \Gamma^M \del_M -M \right] \Psi(x,y) \right\},
}
where $x^\mu\,(\mu = 0,1,2,3)$ are the coordinates of the 4d Minkowski spacetime and $y$ is that of an extra dimension.
We take the extra dimension to be an interval whose length is $L$.
$\Psi(x,y)$ denotes a four-component 5d Dirac spinor with its Dirac conjugate {$\overline{\Psi}$} {defined} as {$\Psi^\dagger \Gamma^0$}, and $\Gamma^M\, (M=0,1,2,3,y)$ are the four-by-four gamma matrices given by
\al{
\Gamma^M =
\begin{cases}
	\gamma^\mu & M=\mu=0,1,2,3,\\
	i \gamma^5 & M=y,
\end{cases}
}
which satisfy the algebra
\al{
\left\{ \Gamma^M, \Gamma^N \right\} = -2 \eta^{MN} \mathbf{1}_4.
}
Here, the 5d metric $\eta^{MN}$ is chosen as $\eta^{MN} = \text{diag}(-1,1,1,1,1)$.
A 5d Dirac spinor $\Psi(x,y)$ is decomposed into the left-handed component $\Psi_L$ and the right-handed one $\Psi_R$ as $\Psi = \Psi_L + \Psi_R$, where the chiral projectors $P_{L/R}$ working as $\Psi_{L/R} = P_{L/R} \Psi$ are defined by $P_{L/R} = (1 \mp \gamma^5)/2$.

As discussed in~\cite{Fujimoto:2011kf}, all the possible {BC's} at $y=0,L$ in this system are classified by use of the action principle, where the following four types are possible:
\al{
&\text{type (I):}\ \ \    \Psi_R(x,0) = \Psi_R(x,L) = 0, \notag \\
&\text{type (II):}\ \     \Psi_L(x,0) = \Psi_L(x,L) = 0, \notag \\
&\text{type (III):}\      \Psi_R(x,0) = \Psi_L(x,L) = 0, \notag \\
&\text{type (IV):}\       \Psi_L(x,0) = \Psi_R(x,L) = 0.
	\label{eq:original_BC's}
}
Also, the action principle gives us the bulk equation of motion, which is just the 5d Dirac equation as $\left[ i \gamma^\mu \del_\mu - \gamma^5 \del_y -M \right] \Psi(x,y) = 0$.
By casting the chiral projectors on it, the equation is decomposed as
\al{
&i \gamma^\mu \del_\mu \Psi_L(x,y) - \D \Psi_R(x,y) = 0, \notag \\
&i \gamma^\mu \del_\mu \Psi_R(x,y) - \D^\dagger \Psi_L(x,y) = 0,
	\label{eq:Dirac_Eq_1}
}
with the two derivative operators $\D \equiv \del_y + M$ and $\D^\dagger \equiv -\del_y + M$.
It is important that the remaining BC's are automatically fixed through the 5d equations as
\al{
&\text{type (I):}\ \ \    \D^\dagger \Psi_L(x,0) =  \D^\dagger \Psi_L(x,L) = 0, \notag \\
&\text{type (II):}\ \     \D \Psi_R(x,0) = \D \Psi_R(x,L) = 0, \notag \\
&\text{type (III):}\      \D^\dagger \Psi_L(x,0) =  \D \Psi_R(x,L) = 0, \notag \\
&\text{type (IV):}\       \D \Psi_R(x,0) = \D^\dagger \Psi_L(x,L) = 0.
}
After the 5d field is Kaluza-Klein~(KK) decomposed as $\Psi(x,y) =  \sum_n \psi_{L,n}(x) f_n(y) + \sum_n \psi_{R,n}(x) g_n(y)$, we can consider particle profiles in terms of mode functions $f_n(y)$ and $g_n(y)$.
{The 4d components obey the 4d Dirac equations},
\al{
i \gamma^\mu \del_\mu \psi_{L,n}(x) - m_n \psi_{R,n}(x) = 0,\quad
i \gamma^\mu \del_\mu \psi_{R,n}(x) - m_n \psi_{L,n}(x) = 0.
	\label{eq:Dirac_Eq_2}
}
From Eqs.~(\ref{eq:Dirac_Eq_1}) and (\ref{eq:Dirac_Eq_2}), we show that {the two} mode functions obey {the} Dirac equations,\footnote{
{Note that these relations are understood through quantum mechanical supersymmetry~\cite{Lim:2005rc,Lim:2007fy,Lim:2008hi,Nagasawa:2008an}.}
}
\al{
\D^\dagger f_n(y) = m_n g_n(y),\quad
\D         g_n(y) = m_n f_n(y),
	\label{eq:Dirac}
}
and also Klein-Gordon equations,
\al{
\D \D^\dagger f_n(y) = m_n^2 f_n(y),\quad
\D^\dagger \D g_n(y) = m_n^2 g_n(y).
	\label{eq:Klein-Gordon}
}
The relation $\D^\dagger \D = \D \D^\dagger = -\del_y^2 + M^2$ is found in the Klein-Gordon operators.
The BC's are represented as conditions on the mode functions by
\al{
&\text{type (I):}\ \ \ 
	\D^\dagger f_n(0) = \D^\dagger f_n(L) = g_n(0) = g_n(L) = 0, \notag \\
&\text{type (II):}\ \ 
	f_n(0) = f_n(L) = \D g_n(0) = \D g_n(L) = 0, \notag \\
&\text{type (III):}\ 
	\D^\dagger f_n(0) = f_n(L) = g_n(0) = \D g_n(L) = 0, \notag \\
&\text{type (IV):}\ 
	f_n(0) = \D^\dagger f_n(L) = \D g_n(0) = g_n(L) = 0.
}

It is not so difficult to solve these quantum mechanical systems and we provide the solutions in the following part.
In every case, {a bound-state solution or a pair of such kind of solutions} ($m_0^2 \leq M^2$) is realizable depending on a value of the bulk mass $M$.
On the other hand, irrespective of a value of $M$, infinite number of positive energy solutions ($m_n^2 > M^2$) are possible, which we usually call KK modes.
Note that the {positive modes} always correspond to Dirac particles, and both of Weyl and Dirac fermions can occur as the bound states.

\begin{itemize}
\item type (I): \\
\vspace{-5mm}
\al{
0)\ 
& m_0^2 = 0, \notag \\
& f_0(y) = \sqrt{\frac{2M}{e^{2ML}-1}} \, e^{My},\quad
  g_0(y): \text{no solution}, \\
n)\ 
& m_n^2 = M^2 + \left( \frac{n\pi}{L} \right)^2 \quad (n=1,2,3,\cdots), \\
& f_n(y) = \frac{1}{\sqrt{2L}} \left( e^{i \frac{n\pi}{L} y} - \frac{M - i\frac{n\pi}{L}}{M + i\frac{n\pi}{L}} e^{-i \frac{n\pi}{L} y} \right), \\
& g_n(y) = {\frac{i}{m_n} \sqrt{\frac{2}{L}}} {\left( M - i\frac{n\pi}{L} \right)} \sin\left( \frac{n\pi}{L} y \right).
}
\item type (II): \\
\vspace{-5mm}
\al{
0)\ 
& m_0^2 = 0, \notag \\
& f_0(y): \text{no solution},\quad
  g_0(y) = \sqrt{\frac{2M}{1-e^{-2ML}}} \, e^{-My}, \\
n)\ 
& m_n^2 = M^2 + \left( \frac{n\pi}{L} \right)^2 \quad (n=1,2,3,\cdots), \\
& f_n(y) = \sqrt{\frac{2}{L}} \sin\left( \frac{n\pi}{L} y \right), \\
& g_n(y) = \frac{1}{m_n} \sqrt{\frac{2}{L}} \left( -\frac{n\pi}{L}
           \cos\left( \frac{n\pi}{L} y \right) +
           M \sin\left( \frac{n\pi}{L} y \right) \right).
}
\item type (III): \\
\vspace{-5mm}
\al{
0)\ 
& m_0^2 = M^2 - \kappa^2 \text{ with } \frac{\kappa}{M} = - \tanh(\kappa L), \\
& f_0(y) =
  \begin{cases}
  \displaystyle
  \sqrt{\frac{\kappa}{\sinh(2 \kappa L) - 2 \kappa L}}
  \left( e^{\kappa(y-L)} - e^{-\kappa(y-L)} \right) & \text{for } ML < -1, \\
  \text{ no solution} & \text{for } ML \geq -1,
  \end{cases} \label{eq:condition_example} \\
&  g_0(y) =
  \begin{cases}
  \displaystyle
  \sqrt{\frac{\kappa}{\sinh(2 \kappa L) - 2 \kappa L}} \frac{2 M}{m_0 \left( e^{\kappa L} + e^{-\kappa L} \right)}
  \left( e^{\kappa y} - e^{-\kappa y} \right) & \text{for } ML < -1, \\
  \text{ no solution} & \text{for } ML \geq -1,
  \end{cases} \\
n)\ 
& m_n^2 = M^2 + k_n^2 \text{ with } \frac{k_n}{M} = -\tan(k_n L) \quad (n={1,2,3,\cdots}), \\
&{k_{n} L  =
\begin{cases}
  \left(n-\frac{1}{2}\right) \pi < k_n L < n \pi & (\text{for } ML>0), \\
  (n-1) \pi < k_n L < \left( n - \frac{1}{2} \right) \pi & (\text{for } -1<ML<0), \\
   n \pi < k_n L < \left(n+\frac{1}{2}\right) {\pi} & (\text{for } ML \leq -1), \\
\end{cases}} \\
& f_n(y) = \sqrt{\frac{1}{\frac{L}{2} - \frac{1}{4k_n} \sin(2 k_n L)}}
           \sin\left( k_n (y-L) \right), \\
& g_n(y) = \frac{1}{m_n} \sqrt{\frac{1}{\frac{L}{2} - \frac{1}{4k_n} \sin(2 k_n L)}}
           \left( -k_n \cos(k_n (y-L)) + M \sin(k_n (y-L)) \right){.}
}
\item type (IV): \\
\vspace{-5mm}
\al{
0)\ 
& m_0^2 = M^2 - \kappa^2 \text{ with } \frac{\kappa}{M} = + \tanh(\kappa L), \\
& f_0(y) =
  \begin{cases}
  \displaystyle
  \sqrt{\frac{\kappa}{\sinh(2 \kappa L) - 2 \kappa L}}
  \left( e^{\kappa y} - e^{-\kappa y} \right) & \text{for } ML > 1, \\
  \text{ no solution} & \text{for } ML \leq 1,
  \end{cases} \\
&  g_0(y) =
  \begin{cases}
  \displaystyle
  \sqrt{\frac{\kappa}{\sinh(2 \kappa L) - 2 \kappa L}} \frac{2 M}{m_0 \left( e^{\kappa L} + e^{-\kappa L} \right)}
  \left( e^{\kappa (y-L)} - e^{-\kappa (y-L)} \right) & \text{for } ML > 1, \\
  \text{ no solution} & \text{for } ML \leq 1,
  \end{cases} \\
n)\ 
& m_n^2 = M^2 + k_n^2 \text{ with } \frac{k_n}{M} = +\tan(k_n L) \quad (n={1,2,3,\cdots}), \\
&{k_{n} L =
\begin{cases}
  \left(n-\frac{1}{2}\right) \pi < k_n L < n \pi & (\text{for } ML<0), \\
  (n-1) \pi < k_n L < \left( n - \frac{1}{2} \right) \pi & (\text{for } 0<ML<1), \\
   n \pi < k_n L < \left(n+\frac{1}{2}\right) {\pi} & (\text{for } ML \geq 1), \\
\end{cases}} \\
& f_n(y) = \sqrt{\frac{1}{\frac{L}{2} - \frac{1}{4k_n} \sin(2 k_n L)}}
           \sin\left( k_n y \right), \\
& g_n(y) = \frac{1}{m_n} \sqrt{\frac{1}{\frac{L}{2} - \frac{1}{4k_n} \sin(2 k_n L)}}
           \left( -k_n \cos(k_n y) + M \sin(k_n y) \right){.}
}
\end{itemize}

Situations are very different between types (I), (II) and types (III), (IV).
In the former category, the lowest energy state is chiral and then massless ($m_0 = 0$), whose chirality is determined by the BC's.
Concretely, a left-handed/right-handed Weyl fermion is realized when we choose the type (I)/(II) BC's.
The bulk mass $M$ makes the profiles localized toward either of the end points and its direction is dictated by the sign of $M$.
After we switch on Yukawa interactions, these fermions form Dirac masses and become massive.
The localized profiles can help us to generate the observed fermion mass hierarchy.

In the latter category, on the other hand, even the lowest mode is Dirac and both of left-handed and right-handed fermions emerge.
In general, the corresponding mass eigenvalue is not zero ($m_0 \not= 0$).
The existence of the Dirac mode depends on not only the type of BC's, but also the value of $ML$.
We should solve the transcendental equations to know exact spectrum, while {the conditions required for consistent solutions, e.g., $ML < -1$ in Eq.~(\ref{eq:condition_example}), are} easy to be derived.
First, we focus on the type (IV), where the condition is that $M$ is positive and $ML$ is greater than one.
The transcendental form $\kappa/M = \tanh{\kappa L}$ is approximated with good precision when $e^{\kappa L} \gg 1$ as
\al{
\kappa = M \tanh(\kappa L) \simeq M \left( 1 - 2 e^{-2 \kappa L} \right),
}
where such a situation is easily achievable by a positive $\kappa$ with $\kappa L \gtrsim \mathcal{O}(1)$.
{Here, we can find the relation when $\kappa L \gtrsim 2 \sim 3$
\al{
\kappa \simeq M \quad (\text{when } \kappa L \gtrsim 2 \sim 3).
	\label{eq:approx_mass_relation}
}}
Now, the corresponding mass eigenvalue $m_0$ is evaluated {semi-analytically} as
\al{
m_0^2 = M^2 - \kappa^2 \simeq 4 M^2 e^{-2 \kappa L} \ {\simeq 4 M^2 e^{-2 M L}.}
	\label{eq:original_m0}
}

Interestingly, we can obtain an exponentially suppressed Dirac mass via the interrelation between the bulk mass and the BC's.
The above formula will be used to generate minuscule active neutrino masses.
The left-handed and right-handed modes are tightly localized around the branes at $y=L$ and $y=0$, respectively for minimizing their overlap.
A significant feature is that the profiles have zero probabilities on either of the branes, which should be required by the BC's.
Concretely speaking, the mode function of the left-handed fermion is zero at $y=0$ ($f_0(0) = 0$), while the right-handed counterpart is zero at $y=L$ ($g_0(L) = 0$). 
This property is fascinating when we try to apply this mechanism to the neutrino sector of the SM.

Finally we touch the situation in the type (III) BC's.
{The major difference is only} in the way of fermion localizations, where the left-handed and right-handed  modes are located around $y=0$ and $y=L$, respectively.
The feature of the lowest mass eigenvalue is the same.
We easily recognize this point after rewriting the bulk mass $M$, which should be negative and $ML < -1$ for realizing a nontrivial solution, as $M = - |M|$.

\section{Implementation
\label{sec:implementation}}

Based on the discussion in the previous section, we try to implement the mechanism to the neutrino sector of the SM.
In a minimal extension of the SM with neutrino Dirac mass terms, we should introduce right-handed $SU(2)_L$ singlet neutrinos and tiny Yukawa couplings should be arranged by hand.
Our mechanism would resolve these unnatural points, where right-handed components are also supplied from 5D $SU(2)_L$ doublets and minuscule active neutrino masses are generated by the dynamics of the extra dimension as we showed before.

This strategy could {look} fine, but one would worry about {the constraint from the invisible decay width of the $Z$ boson} since additional $SU(2)_L$ {non-singlet} right-handed fermions appear in theory {and extra contributions to the invisible channel are severely restricted~\cite{Barate:1999ce,Acciarri:2000ai,Abreu:2000mh,Abbiendi:2000hu,Collaborations:2000aa,ALEPH:2002aa}}.
This problem is hard to be avoided when gauge bosons live in the bulk.
Nevertheless, we can find a way for evading this difficulty when we remember the property that the right-handed components have zero profiles on either of the two branes.
Hereafter, we choose the type (IV) BC's for discussions, where the profile of the (lightest) right-handed mode vanishes on the brane located at $y=L$ ($g_0(L) = 0$).
If all the gauge bosons are completely confined and localized on this brane, this right-handed mode cannot have gauge interactions and {the issue on the invisible channel} is automatically solved.
Note that the corresponding left-handed parts are localized around the brane and thereby interact with the gauge bosons.
In the following part, we make a concrete discussion.

The 5d action of our phenomenological model is as follows:
\al{
S &= S_{\text{EW}} + S_{\text{lepton}}, \notag \\
S_{\text{EW}} &=
	\int \!\! d^4x \int_0^L \!\!\! dy \, \delta(y-L)
	\left\{
	- \frac{1}{4} \sum_{a=1}^3 W^{a}_{\mu\nu} W^{a\mu\nu}
	- \frac{1}{4} B_{\mu\nu} B^{\mu\nu}
	+ H^\dagger (D_\mu D^\mu - M_H^2) H - \frac{\lambda}{4} \left(H^\dagger H\right)^2
	\right\}, \\
S_{\text{lepton}} &=
	\int \!\! d^4x \int_0^L \!\!\! dy \,
	\Bigg\{
	\sum_{i=1}^{3} \left[ \overline{L}_i (i \Gamma^M \del_M - M_{L_i}) L_i \right] +
	\sum_{i=1}^{3} \left[ \overline{E}_i (i \Gamma^M \del_M - M_{E_i}) E_i \right] \notag \\
&\quad
	+ \delta(y-L) \left[ \sum_{i=1}^{3} \zeta_{L_i} \overline{L}_i (i \gamma^\mu D_\mu P_L) L_i
	+ \sum_{i=1}^{3} \zeta_{E_i} \overline{E}_i (i \gamma^\mu D_\mu P_R) E_i
	- \left( \sum_{i,j=1}^{3} \mathcal{Y}_{ij} \overline{L}_i H E_j + \text{h.c.} \right)
	\right] \Bigg\},
	\label{eq:S_lepton}
}
where we only consider the electroweak part ($S_{\text{EW}}$) and the lepton part ($S_{\text{lepton}}$).
The structure of the electroweak part is completely the same as in the SM, {except that} they are located on the brane at $y=L$.
$W_{\mu\nu}^a \, (a=1,2,3)$, $B_{\mu\nu}$, and $H$ stand for the 4d field strength of the $SU(2)_L$ and $U(1)_Y$ gauge {bosons}, and the 4d Higgs doublet, respectively.
The Higgs potential is described by the two parameters $M_H^2$ and $\lambda$.
In this scenario, the property of the Higgs boson is completely the same as it is in the SM.
We require that the parameter $M_H^2$ is negative, which generates spontaneous electroweak symmetry breaking and the $W$ and $Z$ gauge bosons obtain {masses}.
Here, as the SM, $M_H^2$ should be set as the electroweak scale by hand, and then the hierarchy problem cannot be solved.\footnote{
{We note that our strategy on the neutrino mass via boundary conditions would be viable on the RS warped background~\cite{Randall:1999ee}, where the gauge hierarchy problem can be (classically) solved when the Higgs doublet is localized around the TeV brane.}
}

On the other hand, we assume that three $SU(2)_L$ doublet leptons and three $SU(2)_L$ singlet charged leptons live in the bulk and interact with the gauge bosons and the Higgs through the brane-local interactions.
The three types of coefficients $\zeta_{L_i}$, $\zeta_{E_i}$ and $\mathcal{Y}_{ij}$ {have} the mass dimension $-1$.
$D_\mu$ represents the corresponding covariant derivatives.

The brane-local gauge interactions contain kinetic terms and then the existence of them changes the equation of motions and BC's.
It is important to note that the existence of the brane-local kinetic terms does not change the original BC's in Eq.~(\ref{eq:original_BC's}).
Meanwhile, the equation of motions, and also ``derived'' BC's by use of them subsequently, are manifestly deformed by the presence.

{We mention that the situation in the neutrino mass is similar to that in the Higgsless model~\cite{Csaki:2003dt}, where no Higgs doublet is introduced and mass hierarchies and mixings are realized by boundary conditions and/or interactions with brane-local fields.
In our scenario, the Higgs doublet is involved for Yukawa interactions of the SM fermions except for the neutrinos.
Here, we take all the mass parameters are around the 4d Planck scale, where all the KK-excited states are located far above the reach of the LHC and future collider experiments.
It is noted that under the existence of the Higgs doublet, physical masses of the KK particles need not be around a TeV scale for unitarizing the scattering amplitudes of the longitudinal components of the SM gauge bosons.
As widely known, the Higgs doublet maintains the unitarity in the simplest way.
}

We comment on the gauge symmetry on the system.
The gauge symmetry is not exact in our phenomenological description where the fermions couple to the gauge bosons only at the brane and they fly in the bulk.
At the brane, the fermions are transformed {as} 4d gauge rotations, while no such kind of transformation is defined in the bulk since we assume that the fermions are free in this space.
This discontinuity leads to the violation of the gauge symmetries in the system, and subsequently results in the remnant through the fermionic triangle loop diagrams associated with chiral anomalies.
Here, the remnant part should be very small since deviations in effective gauge couplings are strictly restricted.
We quantify the deviations and discuss the condition for keeping the magnitude of them within acceptable ranges in a later part.

\subsection{$SU(2)_L$ doublet part}

\subsubsection{Deformation via brane-local kinetic terms}

At first, we try to look at the $SU(2)_L$ doublet part.
Each of $L_i$ is decomposed as $(\nu_i, e_i)^\text{T}$ with the 5d neutrino field $\nu_i$ and the 5d charged lepton field $e_i$.
To keep the $SU(2)_L$ gauge structure, we assign the same BC's on them.
We choose the type (IV) BC's as the original BC's, where the lightest right-handed fields $\nu_{R,\,i}^{(0)}$ and $e_{R,\,i}^{(0)}$ cannot have gauge interactions on the brane.
Now, the Dirac equations in Eq.~(\ref{eq:Dirac}) {are} modified as
\al{
\D^\dagger f_{n,\,L_i}(y) &= m_{n,\,L_i} g_{n,\,L_i}(y),
	\label{eq:mod_Dirac_1} \\
\D         g_{n,\,L_i}(y) &= m_{n,\,L_i} \left[ 1 + \zeta_{L_i} \delta(y-L) \right] f_{n,\,L_i}(y),
	\label{eq:mod_Dirac_2}
}
where the second equation contains the contribution from the brane-local kinetic term in Eq.~(\ref{eq:S_lepton}).
We adopt the method for treating the localized terms discussed in Refs.~\cite{Csaki:2003sh,Csaki:2005vy}.\footnote{
{Recently, a detailed discussion on a scalar field coupled to a brane on $S^1$ was made in Ref.~\cite{Donini:2015ejd}.}
}

The way of this approach is as follows.
First, we consider that the localized terms are away from the boundary at a distance $\varepsilon$,
which suggest the presence of the localized terms with a Dirac $\delta$-function in the bulk equation of motion.
Next, we put the ``original'' BC's on the fields at the exact position of the corresponding boundary ($y=L$).
The effective BC including the effect of the brane-local terms {can be evaluated by integrating} the bulk equation in Eq.~(\ref{eq:mod_Dirac_2}) after the following manipulation as
\al{
\D         g_{n,\,L_i}(y) = m_{n,\,L_i} \left[ 1 + \zeta_{L_i} \delta(y-(L-\varepsilon)) \right] f_{n,\,L_i}(y),
}
among $y$ within the range of $[L-\varepsilon,\,L]$.
The resultant is obtained by
\al{
\int_{L-\varepsilon}^{L} dy \, \D g_{n,\,L_i}(y) = g_{n,\,L_i}(L) - g_{n,\,L_i}(L-\varepsilon) = - g_{n,\,L_i}(L-\varepsilon) = \zeta_{L_i} m_{n,\,L_i} f_{n,\,L_i}(L-\varepsilon),
}
where we use the original BC at $y=L$ ($g_{n,\,L_i}(L) = 0$).
Finally, we take the limit $\varepsilon \to 0$ and we obtain the following form,
\al{
\zeta_{L_i} m_{n,\,L_i} f_{n,\,L_i}(L) + g_{n,\,L_i}(L) = 0,
	\label{eq:effective_BC_1}
}
where apparently the original BC is recovered in the limit $\zeta_{L_i} \to 0$.

After this, we focus on the bound-state solution ($n=0$).
From the original BC's at $y=0$ (with Eq.~(\ref{eq:mod_Dirac_1})), the forms of $f_{0,\,L_i}$ and $g_{0,\,L_i}$ are partly fixed as
\al{
f_{0,\,L_i}(y) &= \A_{L_i} \left( e^{\kappa_{L_i} y} - e^{-\kappa_{L_i} y} \right),
	\label{eq:f0Li_form} \\
g_{0,\,L_i}(y) &= \frac{\A_{L_i}}{m_{0,\,L_i}} \left( (-\kappa_{L_i} +M_{L_i}) e^{\kappa_{L_i} y} - (\kappa_{L_i} +M_{L_i}) e^{-\kappa_{L_i} y} \right),
}
with a normalization factor $\A_{L_i}$.
Through the equation in (\ref{eq:effective_BC_1}), we can reach the relation
\al{
\zeta_{L_i} m_{0,\,L_i}^2 \tanh(\kappa_{L_i} L) + M_{L_i} \tanh(\kappa_{L_i} L) = \kappa_{L_i},
	\label{eq:mass_relation_withBLKT}
}
which is the deformed condition to determine the physical mass spectrum $m_{0,\,L_i}^2 = M^2_{L_i} - \kappa^2_{L_i}$.

{Here, we should emphasize that our interest is in the case that the value of 
$m_{0,\,L_{i}}$ is extremely small, where such a situation is naturally realized 
by the bulk mass and the original BC's with $\zeta_{L_{i}}=0$
(see Eq. (\ref{eq:original_m0})).
The existence of the brane-local parameter $\zeta_{L_{i}}$ would change the
value of $m_{0,\,L_{i}}$, but the exponential smallness of $m_{0,\,L_{i}}$
should be preserved even with $\zeta_{L_{i}} \ne 0$.
Actually, we find that $m^{2}_{0,\,L_{i}}$ with a nonzero $\zeta_{L_{i}}$ is,
by solving the equation (\ref{eq:mass_relation_withBLKT}), approximately
given by
\al{
m_{0,\,L_i}^2 
\simeq \frac{4 M^2_{L_i} e^{-2 M_{L_i} L}}{1 + 2 \zeta_{L_i} M_{L_i}},
	\label{eq:mass_relation}
}
which is exponentially small with $M_{L_i} L \gtrsim 2 \sim 3$.
Thereby, the modification originated from a nonzero $\zeta_{L_{i}}$ would not 
be so significant for the exponential suppression of $m_{0,\,L_{i}}$.}

The presence of the brane-local part enforces to re-evaluate the normalization factor $\A_{L_i}$ in $f_{0,\,L_i}$ as
\al{
\int_0^L dy \left[ 1 + \zeta_{L_i} \delta(y-L) \right] f_{0,\,L_i}^2 (y) = 1,
}
which leads to
\al{
  \A_{L_i} = \sqrt{\frac{1}{\left(\sinh(2 \kappa_{L_i} L) - 2 \kappa_{L_i} L\right)/\kappa_{L_i}
  + 2 \zeta_{L_i} {\left( \cosh(2 \kappa_{L_i} L) -1 \right)}  }}.
	\label{eq:normalization_factor}
}

At the end of this section, we comment on the value of $g_{0,\,L_i}$ at the boundary $y=L$ after the modification.
Now, the Eq.~(\ref{eq:effective_BC_1}) says,
\al{
g_{0,\,L_i}(L) = - \zeta_{L_i} m_{0,\,L_i} f_{0,\,L_i}(L),
	\label{eq:modified_value_y=L}
}
which is no more zero even though the right-hand side would be very small since $m_{0,\,L_i} L \ll 1$ is required within our interest.
On the other hand, the chirality projector in Eq.~(\ref{eq:S_lepton}) makes the right-handed modes still completely decoupled from the brane-localized gauge bosons.
Thereby, there is still no need for worrying about {additional contributions to the $Z$ boson invisible width}.

\subsubsection{Neutrino mass}

Here, we look at target values of the bulk masses for realizing the observed neutrino mass hierarchy.
{In our scenario, the neutrino masses are given as {tiny masses}, where no 5d {$SU(2)_L$} singlet neutrino is introduced.}
For simplicity, we only focus on the normal hierarchy in the neutrino mass ordering.
{A latest combined result by Bayesian method} is announced in Ref.~\cite{Bergstrom:2015rba} as $\Delta m_{21}^2 = 7.5 \times 10^{-5}\,\text{eV}^2$ and $\Delta m_{32}^2 = 2.457 \times 10^{-3}\,\text{eV}^2$ at the best fit point in the $\chi^2$ analysis.
When we fix $m_1$ as $0.01\,\text{eV}$, the other two are determined as $m_2 \simeq 0.0132\,\text{eV}$ and $m_3 \simeq 0.0498\,\text{eV}$, respectively.
The mass relation in Eq.~(\ref{eq:mass_relation}) is written as
\al{
{m_{0,\,L_i}
\simeq \frac{2 M_{L_i} e^{- M_{L_i} L}}{\sqrt{1 + 2 \zeta_{L_i} M_{L_i}}}
= \frac{2 \widetilde{M}_{L_i} (L^{-1}) e^{- \widetilde{M}_{L_i}}}{\sqrt{1 + 2 \widetilde{\zeta}_{L_i} \widetilde{M}_{L_i}}}},
}
{with dimensionless variable $\widetilde{M}_{L_i} \equiv M_{L_i} L$.
In this analysis, we set the mass scale $L^{-1}$ at the 4d Planck mass $M_{pl} = 1.22 \times 10^{19}\,\text{GeV}$.}
{As we discuss later, the lower bound of $\widetilde{\zeta}_{L_i}$ is estimated as $\widetilde{\zeta}_{L_i} \sim 10$.
Here, we use the value $\widetilde{\zeta}_{L_i} = 10$ for estimation.}
When we adopt the following choice,
\al{
{\widetilde{M}_{L_{1}}} = {70.6},\quad
{\widetilde{M}_{L_{2}}} = {70.3},\quad
{\widetilde{M}_{L_{3}}} = {69.0},
	\label{eq:tildeM_choice}
}
the realized neutrino masses are
\al{
{m_{0,\,L_1}} = m_{\nu_1} \simeq {0.010}\,\text{eV},\quad
{m_{0,\,L_2}} = m_{\nu_2} \simeq {0.013}\,\text{eV},\quad
{m_{0,\,L_3}} = m_{\nu_3} \simeq {0.049}\,\text{eV}.
}
Now, we show that our mechanism works well for generating the order of the minuscule observed neutrino masses with no serious parameter tuning.
Note that a bit parameter tuning would be required when we focus on the observed result with good precision.

\subsubsection{Constraints via gauge coupling deviation}

In the current configuration, the $SU(2)_L$ doublet leptons live both in the bulk and the brane at $y=L$, which forces to re-normalize the wave function profiles of the leptons as concretely calculated in Eq.~(\ref{eq:normalization_factor}).
The contribution from the bulk to the factor produces a deviation in the $SU(2)_L$ gauge coupling {$g$} from the value in the SM as
\al{
g \, \zeta_{L_i} \int_{0}^{L} dy \, \delta(y-L) f_{0,\,L_i}^2(y) &= g
\left(\frac{2 \zeta_{L_i} {\left( \cosh(2 \kappa_{L_i} L) -1 \right)} }{\left(\sinh(2 \kappa_{L_i} L)
  - 2 \kappa_{L_i} L\right)/\kappa_{L_i}
  + 2 \zeta_{L_i} {\left( \cosh(2 \kappa_{L_i} L) -1 \right)} } \right) \notag \\
&\simeq
g \left(\frac{2 \widetilde{\zeta}_{L_i}}{1/\widetilde{\kappa}_{L_i} + 2 \widetilde{\zeta}_{L_i}}\right) \notag\\
&\equiv g (1 + a_i),
	\label{eq:doublet_SU(2)L_coupling_deviation}
}
where we assume that $\kappa_{L_i} L$ is not {small}.
Note that this form has a dependence on the generation shown by the index $i$.
The form of the deviation in the $U(1)_Y$ gauge interaction takes the same.

This type of deviation is severely constrained by electroweak precision measurements.
The calculation of the Fermi constant $G_F$ yields
\al{
G_F = \frac{(g(1+ a_i))^2}{4\sqrt{2} m_W^2}
= \frac{g^2}{4\sqrt{2} m_W^2}
+ \frac{g^2 (2a_i + a^2_i)}{4\sqrt{2} m_W^2}
\equiv G_{F,\,0} + \delta G_F.
}
We estimate the bound through the Peskin-Takeuchi $S,\,T,\,U$ parameters~\cite{Peskin:1990zt,Peskin:1991sw}, which are related to the deviation in the Fermi constant $\delta G_F$ as~\cite{Rizzo:1999br,Davoudiasl:1999tf,Csaki:2002gy,Flacke:2011nb}
\al{
S = 0,\quad
T = - \frac{1}{\alpha} \frac{\delta G_F}{G_F},\quad
U = \frac{4 \sin^2{\theta_W}}{\alpha} \frac{\delta G_F}{G_F},
}
where $\alpha$ is the electromagnetic fine structure constant and $\sin{\theta_W}$ is the sine of the  Weinberg angle in the $\overline{MS}$ scheme, both given at the scale $m_Z$ as $\alpha(m_Z) \simeq 1/127.916$, $\sin^2{\theta_W} \simeq 0.2313$, respectively~\cite{Agashe:2014kda}.
The factor $\delta G_F/G_F$ is easy to be estimated as
\al{
\frac{\delta G_F}{G_F} = \frac{2a_i + a_i^2}{(1+a_i)^2} \simeq 2a_i,
}
since $a_i$ should be $|a_i| \ll 1$.
The latest values of the oblique parameters reported by the {\it Gfitter} group~\cite{Baak:2014ora} are
$S=0.05\pm0.11$, $T=0.09\pm0.13$, $U=0.01\pm0.11$ in the reference point $m_{t,\,\text{ref}} = 173\,\text{GeV}$ and $m_{h,\,\text{ref}} = 125\,\text{GeV}$.
The correlation coefficients between the three parameters are given by $\rho_{ST} = +0.90$, $\rho_{SU} = -0.59$, $\rho_{TU} = -0.83$, respectively.

To perform a $\chi^2$ analysis gives us the allowed region of $a_i$ with a $95\%$ confidence level as
\al{
-{7.62} \times 10^{-4} \lesssim a_i \lesssim {1.99} \times 10^{-4}.
	\label{eq:allowed_region_GF}
}
Note that the factor $a_i$ tends to be negative in our case and we focus on the lower bound.
Following the discussion in the previous section, the parameters $\widetilde{\kappa}_{L_i}\,(i=1,2,3)$ should be {around $70$}.
When we fix $\widetilde{\kappa}_{L_i} {\simeq \widetilde{M}_{L_i}} = 70$, the brane-local parameters $\widetilde{\zeta}_{L_i}$ should fulfill the condition
\al{
\widetilde{\zeta}_{L_i} \gtrsim {9.4},
	\label{eq:condition_via_GF}
}
to circumvent the bound.

We add a few comments.
The lepton universality is not severely violated if the condition in Eq.~(\ref{eq:condition_via_GF}) is realized.
The deviation in the $SU(2)_L$ gauge coupling also modifies the tree level unitarity condition for longitudinal components of the massive gauge bosons.

{On the other hand, processes with lepton flavor violation are tightly constrained by experiments.
In the following part, we concretely have a discussion on the bound via the $Z$-boson related processes, $Z \to \mu^{\pm} e^{\mp}$ and $\mu^{-} \to e^{-} Z^{\ast} \to e^{-} e^{+} e^{-}$ (${}^{\ast}$ implying offshellness of the intermediate particle).}
{When $a_{i}$ is not universal among $i = 1,2,3$, the lepton flavor violating part emerges in the vertex {$\overline{e'}_{L} \gamma^{\mu} Z_{\mu} \mu'_{L}$}, where the leptons in their mass eigenstates are designated with the prime symbol.
In the present scenario, as we explicitly mention later in Eq.~(\ref{eq:UPMNS_relation}), the Yukawa couplings of the neutrinos are diagonal and then the left-handed charged lepton fields should be transformed as
\al{
e_{L,\,i} = (U_{\text{PMNS}}^{\dagger})_{ij} \, e'_{L,\,j}.
}
{Here, we assume that the lepton Yukawa matrix is diagonalized only by the unitary transformation for the left-handed charged leptons, without nontrivial unitary transformation for the right-handed ones.}
In this circumstance, lepton flavor violation occurs only in $\overline{e'}_{L} \gamma^{\mu} Z_{\mu} \mu'_{L}$.
The coefficient of this operator {$C_{\overline{e'}_{L} Z \mu'_{L}}$} is easily {calculated with the notation of~\cite{Denner:1991kt}} as
\al{
{C_{\overline{e'}_{L} Z \mu'_{L}} \equiv g_{L, \ell}^{Z} \, \delta g,\quad
g_{L, \ell}^{Z} = e \left[ \frac{I^{3}_{{W,\ell}} - s_{W}^{2} Q_{\ell}}{s_{W} c_{W}} \right], \quad
\delta g \equiv
\sum_{i=1}^{3} (U_{\text{PMNS}})_{1, i} \, (U_{\text{PMNS}}^{\ast})_{2, i} \, a_{i},}
	\label{eq:deltag}
}
{where $e$, $I^{3}_{{W,\ell}}$ and $Q_{\ell}$ stand for the electromagnetic charge of the positron, the weak isospin of the charged lepton ($\ell$) and the electromagnetic charge of $\ell$ in the unit of $e$, respectively.
Also, we adopt the short-hand notations, $s_{W} = \sin{\theta_{W}}$ and $c_{W} = \cos{\theta_{W}}$.}
We used the unitary condition $(U_{\text{PMNS}}) (U_{\text{PMNS}}^{\dagger}) = 1_{3}$, which suggests that
$\delta g$ goes to zero if $a_{1} = a_{2} = a_{3}$\,(universal case).
Here, we adopt the standard notation on $U_{\text{PMNS}}$~\cite{Agashe:2014kda} as
\al{
U_{\text{PMNS}}
=
\begin{pmatrix}
c_{12} c_{13} &
s_{12} c_{13} &
s_{13} e^{- i \delta_{\text{CP}}} \\
- s_{12} c_{23} - c_{12} s_{23} s_{13} e^{i \delta_{\text{CP}}} &
c_{12} c_{23} - s_{12} s_{23} s_{13} e^{i \delta_{\text{CP}}} &
s_{23} c_{13} \\
s_{12} s_{23} - c_{12} c_{23} s_{13} e^{i \delta_{\text{CP}}} &
- c_{12} s_{23} - s_{12} c_{23} s_{13} e^{i \delta_{\text{CP}}} &
c_{23} c_{13}
\end{pmatrix},
}
with setting the two Majorana CP angles as zero since our neutrino mass matrix is Dirac-type.
We adopt the following digits for our estimation, $s_{12}^{2} = 0.304$, $s_{13}^{2} = 0.0218$, $s_{23}^{2} = 0.452$ reported in~\cite{Bergstrom:2015rba} as best fit values of a global analysis {in the case of the normal mass ordering}.

\begin{figure}[t]
\begin{center}
\includegraphics[width=0.6\columnwidth]{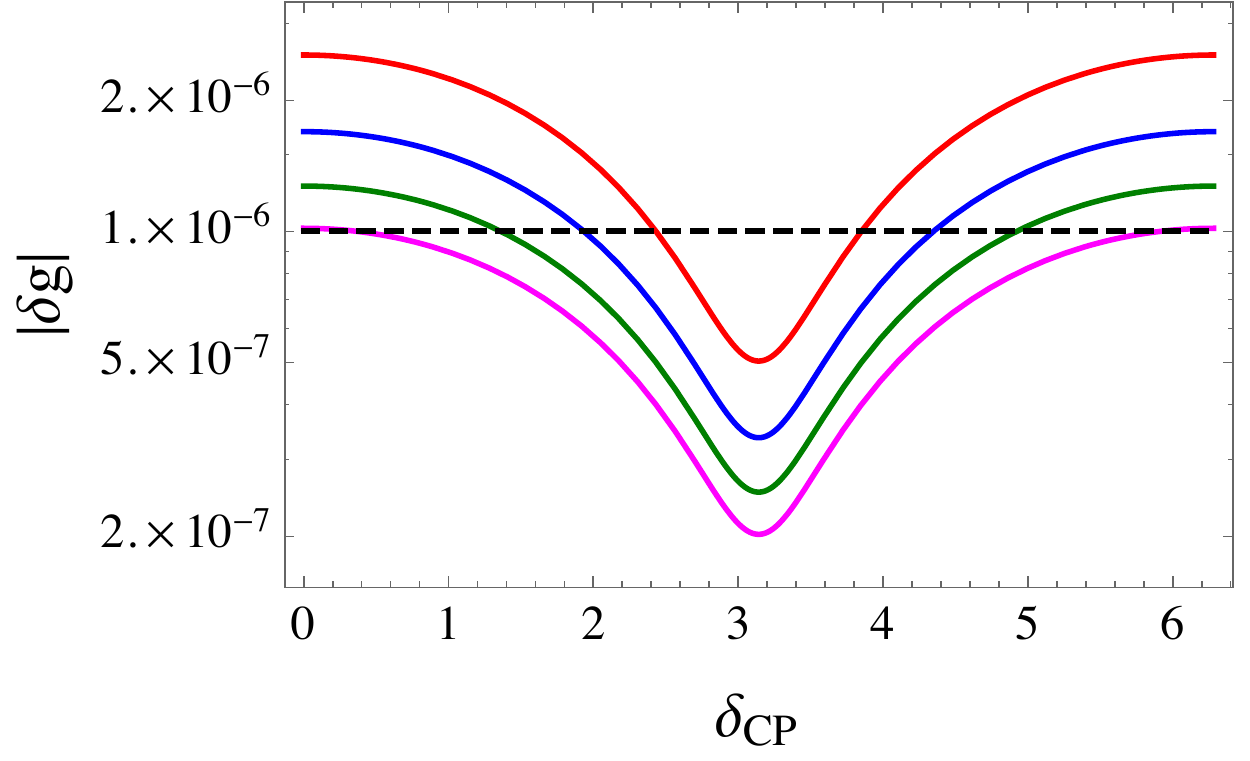}
\caption{
{Distributions of $|\delta g|$ [defined in Eq.~(\ref{eq:deltag})] as functions of $\delta_{\text{CP}}$ with $\widetilde{M}_{L_{1}} = 70.6$, $\widetilde{M}_{L_{2}} = 70.3$, $\widetilde{M}_{L_{3}} = 69.0$ as adopted in Eq.~(\ref{eq:tildeM_choice}), where the red, blue, green, magenta curves correspond to the choices of $\widetilde{\zeta}_{L_{i}}$\,(universal value) $=10$, $15$, $20$, $25$, respectively.
The horizontal dashed line indicates a typical upper bound on $|\delta g|$ via the null observation of the lepton flavor violating process $\mu^{-} \to e^{-} e^{+} e^{-}$ in experiments with a $95\%$ confidence level.}
}
\label{Fig:deltag_curve}
\end{center}
\end{figure}

The upper bound $\text{Br}(\mu^{-} \to e^{-} e^{+} e^{-}) < 1.0 \times 10^{-12}$\,($90\%$ confidence level)~\cite{Agashe:2014kda} puts a  constraint on $\delta g$ as
\al{
|\delta g| \lesssim 10^{-6}. 
}
{Here, we do not take account of the multiplicative factor of a few originating from the difference between the $W$ and $Z$-related gauge interactions.}\footnote{
{We list the concrete digits, $g_{L, \ell}^{Z} \simeq -0.64\,e$ and the $W$-boson counterpart $g_{L, \ell}^{W} = e/(\sqrt{2} s_{W}) \simeq 1.5\,e$, which appears in the dominant decay channel of $\mu^{-}$, $\mu^{-} \to \nu_{\mu} (W^{-})^{\ast} \to e^{-} \nu_{\mu} \overline{\nu_{e}}$.
Thus, we do not underestimate the bound on $\delta g$.}
}
We note that the bound on $\delta g$ from $\text{Br}(Z \to e^{\pm} \mu^{\mp}) < 1.6 \times 10^{-6}$\,($95\%$ confidence level)~\cite{Agashe:2014kda} is much weaker, and thus we can ignore it.
We plot the distributions of $|\delta g|$ when we adopt the values $\widetilde{M}_{L_{1}} = 70.6$, $\widetilde{M}_{L_{2}} = 70.3$, $\widetilde{M}_{L_{3}} = 69.0$ to derive typical neutrino mass scales as discussed around Eq.~(\ref{eq:tildeM_choice}) as functions of the Dirac CP phase $\delta_{\text{CP}}$ in $U_{\text{PMNS}}$.
Here, the red, blue, green, magenta curves correspond to the choices of $\widetilde{\zeta}_{L_{i}}$\,(universal value) $=10$, $15$, $20$, $25$, respectively.
The horizontal dashed line indicates a typical upper bound on $|\delta g|$ via the null observation of the lepton flavor violating process $\mu^{-} \to e^{-} e^{+} e^{-}$ in experiments with a $95\%$ confidence level.
From Fig.~\ref{Fig:deltag_curve}, we immediately recognize that when $\widetilde{\zeta}_{L_{i}} \gtrsim 25$, the bound on $|\delta g|$ is evaded irrespective of the value of $\delta_{\text{CP}}$.
We mention that this bound is tighter than that via the $S$ and $T$ parameters in Eq.~(\ref{eq:condition_via_GF}), while the difference is not so significant.
}

Only little room in the deviation of the gauge coupling from the SM is, as shown in Eq.~(\ref{eq:allowed_region_GF}), and then no tight constraint comes from this phenomenon.\footnote{
{See Refs.~\cite{SekharChivukula:2001hz,Abe:2003vg,Chivukula:2003kq,Csaki:2003dt,Ohl:2003dp,Abe:2004wv,Sakai:2006qi,Nishiwaki:2010te} for unitarity in models on extra dimensions.}
}

\subsection{$SU(2)_L$ singlet part}

\subsubsection{Deformation via brane-local kinetic terms}

For the right-handed components of the charged leptons, we should arrange the type (II) BC's for realizing localized right modes.
Like the neutrino case in the previous section, we can realize the mass hierarchy by the help of the bulk masses.
From Eq.~(\ref{eq:S_lepton}), the Dirac equation is given by
\al{
\D^\dagger f_{n,\,E_i}(y) &= m_{n,\,E_i} \left[ 1 + \zeta_{E_i} \delta(y-L) \right] g_{n,\,E_i}(y),
	\label{eq:mod_Dirac_3} \\
\D         g_{n,\,E_i}(y) &= m_{n,\,E_i} f_{n,\,E_i}(y),
	\label{eq:mod_Dirac_4}
}
and following the method applied in the doublet case leads to the {mass-determining} condition,
\al{
m_{n,\,E_i} \zeta_{E_i} g_{n,\,E_i}(L) - f_{n,\,E_i}(L) = 0.
}
We can recognize that the BC at $y=L$ for KK modes $(n \not= 0)$ is deformed as above, while
the BC at $y=L$ for the massless bound state $(n=0,\ m_{0,\,E_i} = 0)$ is intact as $f_{n,\,E_i}(L) = 0$.
Then, the BC for $g_{n,\,E_i}(y)$ at $y=L$ through the equation of motion in Eq.~(\ref{eq:mod_Dirac_4}) is also intact and the original right-handed massless zero mode can exist under the presence of a nonzero $\zeta_{E_i}$.
This is because the brane-local kinetic term holds right chirality and a massless particle is still massless under the re-normalization of the kinetic term.
Here, we consider that the bulk masses {$M_{E_i}$} are negative {($M_{E_i} = - |M_{E_i}|$)} to make a sizable difference at $y=L$ for explaining the mass hierarchy in the charged leptons through Yukawa interactions.
The normalization factor of the zero modes is modified as
\al{
\int_0^L dy \left[ 1 + \zeta_{E_i} \delta(y-L) \right] g_{0,\,E_i}^2 (y) = 1,
}
which means
\al{
g_{0,\,E_i}(y) &= \A_{E_i} e^{|{M_{E_i}}| y}, 
	\label{eq:g0Ei_form} \\
\A_{E_i} &= \sqrt{\frac{2 |M_{E_i}|}{(1 + 2 |M_{E_i}| \zeta_{E_i}) e^{2 |{M_{E_i}}| L} -1}}.
}
The deviation in the $U(1)_Y$ gauge coupling, {where $g'$ is the value in the SM,} is estimated as
\al{
g' \zeta_{E_i} \int_{0}^{L} dy \, \delta(y-L) g_{0,\,E_i}^2(y) &= g'
\left(
\frac{2 |M_{E_i}| \zeta_{E_i} e^{2 |{M_{E_i}}| L}}{(1 + 2 |M_{E_i}| \zeta_{E_i}) e^{2 |{M_{E_i}}| L} -1}
\right) \notag \\
&\simeq
g' \left(
\frac{2 |M_{E_i}| \zeta_{E_i}}{(1 + 2 |M_{E_i}| \zeta_{E_i})}
\right).
}
Like the doublet case, if the {dimensionless} factor $|M_{E_i}| \zeta_{E_i}$ is quite large compared with unity as
\al{
|M_{E_i}| \zeta_{E_i} \gg 1,
}
the magnitude of the deviations can be within acceptable ranges.\footnote{
We note that the left-hand components of the charged leptons via $SU(2)_L$ doublets also possess $U(1)_Y$ charges and corresponding effective gauge couplings deviate from the SM.
The magnitude of the deviations {is} easily estimated by the replacement $g \to g'$ in Eq.~(\ref{eq:doublet_SU(2)L_coupling_deviation}).}

Here, we briefly mention about the quark sector.
When we assign the type (I) BC's for quark doublets and type (II) BC's for quark singlets, we obtain all the Weyl fermions for describing the quark sector of the SM as zero modes.
Since the matter content is the same as it in the SM, no additional exotic particle contributing to the chiral anomalies emerge.

\subsubsection{Charged lepton mass and lepton mixing structure}

The mass terms for the charged leptons are symbolically written down as
\al{
\overline{e_{L_i}^{(0)}}
\left[
m_\nu e_{R_i}^{(0)} + m_\ell E_{R_i}^{(0)} + \text{h.c.}
\right],
}
where $m_\nu$ and $m_\ell$ are typical scales of the active neutrinos and the charged lepton ($m_\nu \ll m_\ell$), respectively.
Note that $e_{R_i}^{(0)}$ originates from the 5D $SU(2)_L$ doublet $L_i$ and its mode function is the same as that of {$\nu_{R_i}^{(0)}$}, where the profile is (almost) zero, as shown around Eq.~(\ref{eq:modified_value_y=L}), on the brane at $y=L$ where the gauge bosons and the Higgs bosons are localized.
{The components are sterile to the gauge and Higgs bosons}
because of the value of the wavefunction at $y=L$ and the chiral projector in Eq.~(\ref{eq:S_lepton}).
Thereby, we can neglect them in phenomenology {and} the structure of the charged leptons gets to be identical with the SM.
{We mention that the mixing effect between $e_{R_i}^{(0)}$ and $E_{R_i}^{(0)}$ is negligible since the coefficients are hierarchical very much (at least $m_\nu/m_\ell < 10^{-6}$ when $\ell = e$).}

We put a comment on the neutrino mixings.
In our model, the neutrino mass matrix is diagonal, while non-diagonal components are available in the charged lepton Yukawa sectors as shown in Eq.~(\ref{eq:S_lepton}).
Therefore, not only the mass scales of the active neutrinos, but also the mixing patterns including the Dirac CP phase would be achievable when we realize the following condition,
\al{
\widetilde{\mathcal{Y}} \left(\frac{v}{\sqrt{2}}\right) =
(U_{\text{PMNS}})^\dagger
\begin{pmatrix}
m_e & & \\
& m_\mu & \\
& & m_\tau
\end{pmatrix},
	\label{eq:UPMNS_relation}
}
where $\widetilde{\mathcal{Y}}$ represents the three-by-three effective Yukawa matrix for the charged leptons after executing the integral along the $y$ direction, {which is given by}
\al{
{\widetilde{\mathcal{Y}}_{ij} =\mathcal{Y}_{ij}  \int_0^L dy \, (f_{0,\, L_i}(y))^\ast g_{0,\, E_j}(y) \,
\delta(y-L).}
}
$v \simeq 246\,\text{GeV}$ is the Higgs vacuum expectation value.
We can realize this situation by a suitable set of the bulk masses and the brane-local parameters, also adjusting the components of the {three-by-three 5d Yukawa matrix}.
{The exponential forms in $f_{0,\, L_i}$ and $g_{0,\, E_j}$ in Eqs.~(\ref{eq:f0Li_form}) and (\ref{eq:g0Ei_form}) help us to realize the mass hierarchy in the charged leptons naturally.}

\section{Summary and Discussion
\label{eq:conclusion}}

In this paper, we had {discussions} on a new mechanism for generating the minuscule active neutrino masses 
via 5d $SU(2)_L$ lepton doublets via Dirac mass terms without introducing gauge singlet right-handed neutrinos in the model.
Due to the asymmetric BC's for the doublets, the left and right components are localized around {the} boundaries and tiny Dirac masses are naturally realized due to minute overlaps of them.
{This mechanism provides a new tool for building neutrino mass models.}
Also, if the gauge bosons and the Higgs boson are localized on one of the branes, the additional right-handed modes have no interaction with gauge bosons.
In such a situation realized on an interval, we can circumvent the tight bound from the invisible decay width of the $Z$ boson precisely measured by the LEP experiments~\cite{Barate:1999ce,Acciarri:2000ai,Abreu:2000mh,Abbiendi:2000hu,Collaborations:2000aa,ALEPH:2002aa}.

{Finally, we shall see an implication of the proposed model to cosmology.} Big-Bang nucleosynthesis\,(BBN) is a remarkable achievement of the standard Big-Bang cosmology \cite{Agashe:2014kda}. The accuracy of the prediction and observation of BBN has being improved. The results allow constrains on physics beyond the SM. The main constraint comes from the energy density of relativistic degree of freedom at temperature, $\mbox{T} \simeq 1\, \mbox{MeV}$, when BBN was about to begin. The relativistic degree of freedom is often denoted as $g_{\ast}$. In SM case the value is
estimated as $g_{\ast}=10.75$. In the present model the exotic fields of $SU(2)_L$ doublets, $(\nu_i,e_i)^{(0)}_{R}\,(i=1,2,3)$, appear, which have no interaction (being sterile) and form Dirac mass {terms} of order $m_{\nu}$. The sterile fields were decoupled from thermal bath at a sufficient early universe, and the chirality-flip production from the active $SU(2)_L$ doublets, $(\nu_i,e_i)^{(0)}_{L}$, through the tiny Dirac mass {terms} are negligible as shown in Ref.\,\cite{Dolgov:2002wy}. The sterile fields do not contribute to the effective degree $g_{\ast}$, and then, the present model is consistent with BBN.

\section*{Acknowledgments}

We are grateful for Hiroyuki~Ishida, Taichiro~Kugo and Yusuke~Shimizu for fruitful discussions.
This work is supported in part by the Grant-in-Aid for Scientific Research (No.15K05055~(M.S.)) by the Japanese Ministry of Education, Science, Sports and Culture.

\bibliographystyle{JHEP}
\bibliography{draft}

\providecommand{\href}[2]{#2}\begingroup\raggedright\begin{thebibliography}{100}

\bibitem{An:2012eh}
{\scshape Daya Bay} collaboration, F.~P. An et~al., \emph{{Observation of
  electron-antineutrino disappearance at Daya Bay}},
  \href{http://dx.doi.org/10.1103/PhysRevLett.108.171803}{\emph{Phys. Rev.
  Lett.} {\bf 108} (2012) 171803}, [\href{http://arxiv.org/abs/1203.1669}{{\tt
  1203.1669}}].

\bibitem{Ahn:2012nd}
{\scshape RENO} collaboration, J.~K. Ahn et~al., \emph{{Observation of Reactor
  Electron Antineutrino Disappearance in the RENO Experiment}},
  \href{http://dx.doi.org/10.1103/PhysRevLett.108.191802}{\emph{Phys. Rev.
  Lett.} {\bf 108} (2012) 191802}, [\href{http://arxiv.org/abs/1204.0626}{{\tt
  1204.0626}}].

\bibitem{Adamson:2013whj}
{\scshape MINOS} collaboration, P.~Adamson et~al., \emph{{Measurement of
  Neutrino and Antineutrino Oscillations Using Beam and Atmospheric Data in
  MINOS}}, \href{http://dx.doi.org/10.1103/PhysRevLett.110.251801}{\emph{Phys.
  Rev. Lett.} {\bf 110} (2013) 251801},
  [\href{http://arxiv.org/abs/1304.6335}{{\tt 1304.6335}}].

\bibitem{Abe:2013hdq}
{\scshape T2K} collaboration, K.~Abe et~al., \emph{{Observation of Electron
  Neutrino Appearance in a Muon Neutrino Beam}},
  \href{http://dx.doi.org/10.1103/PhysRevLett.112.061802}{\emph{Phys. Rev.
  Lett.} {\bf 112} (2014) 061802}, [\href{http://arxiv.org/abs/1311.4750}{{\tt
  1311.4750}}].

\bibitem{Abe:2014lus}
{\scshape Double Chooz} collaboration, Y.~Abe et~al.,
  \emph{{Background-independent measurement of $\theta_{13}$ in Double Chooz}},
  \href{http://dx.doi.org/10.1016/j.physletb.2014.04.045}{\emph{Phys. Lett.}
  {\bf B735} (2014) 51--56}, [\href{http://arxiv.org/abs/1401.5981}{{\tt
  1401.5981}}].

\bibitem{Maki:1962mu}
Z.~Maki, M.~Nakagawa and S.~Sakata, \emph{{Remarks on the unified model of
  elementary particles}},
  \href{http://dx.doi.org/10.1143/PTP.28.870}{\emph{Prog. Theor. Phys.} {\bf
  28} (1962) 870--880}.

\bibitem{Pontecorvo:1967fh}
B.~Pontecorvo, \emph{{Neutrino Experiments and the Problem of Conservation of
  Leptonic Charge}}, {\emph{Sov. Phys. JETP} {\bf 26} (1968) 984--988}.

\bibitem{Tortola:2012te}
D.~Forero, M.~Tortola and J.~Valle, \emph{{Global status of neutrino
  oscillation parameters after Neutrino-2012}},
  \href{http://dx.doi.org/10.1103/PhysRevD.86.073012}{\emph{Phys.Rev.} {\bf
  D86} (2012) 073012}, [\href{http://arxiv.org/abs/1205.4018}{{\tt
  1205.4018}}].

\bibitem{Fogli:2012ua}
G.~Fogli, E.~Lisi, A.~Marrone, D.~Montanino, A.~Palazzo et~al., \emph{{Global
  analysis of neutrino masses, mixings and phases: entering the era of leptonic
  CP violation searches}},
  \href{http://dx.doi.org/10.1103/PhysRevD.86.013012}{\emph{Phys.Rev.} {\bf
  D86} (2012) 013012}, [\href{http://arxiv.org/abs/1205.5254}{{\tt
  1205.5254}}].

\bibitem{GonzalezGarcia:2012sz}
M.~Gonzalez-Garcia, M.~Maltoni, J.~Salvado and T.~Schwetz, \emph{{Global fit to
  three neutrino mixing: critical look at present precision}},
  \href{http://dx.doi.org/10.1007/JHEP12(2012)123}{\emph{JHEP} {\bf 1212}
  (2012) 123}, [\href{http://arxiv.org/abs/1209.3023}{{\tt 1209.3023}}].

\bibitem{Capozzi:2013csa}
F.~Capozzi, G.~Fogli, E.~Lisi, A.~Marrone, D.~Montanino et~al., \emph{{Status
  of three-neutrino oscillation parameters, circa 2013}},
  \href{http://dx.doi.org/10.1103/PhysRevD.89.093018}{\emph{Phys.Rev.} {\bf
  D89} (2014) 093018}, [\href{http://arxiv.org/abs/1312.2878}{{\tt
  1312.2878}}].

\bibitem{Gonzalez-Garcia:2014bfa}
M.~C. Gonzalez-Garcia, M.~Maltoni and T.~Schwetz, \emph{{Updated fit to three
  neutrino mixing: status of leptonic CP violation}},
  \href{http://dx.doi.org/10.1007/JHEP11(2014)052}{\emph{JHEP} {\bf 11} (2014)
  052}, [\href{http://arxiv.org/abs/1409.5439}{{\tt 1409.5439}}].

\bibitem{Bergstrom:2015rba}
J.~Bergstrom, M.~C. Gonzalez-Garcia, M.~Maltoni and T.~Schwetz, \emph{{Bayesian
  global analysis of neutrino oscillation data}},
  \href{http://dx.doi.org/10.1007/JHEP09(2015)200}{\emph{JHEP} {\bf 09} (2015)
  200}, [\href{http://arxiv.org/abs/1507.04366}{{\tt 1507.04366}}].

\bibitem{Gonzalez-Garcia:2015qrr}
M.~C. Gonzalez-Garcia, M.~Maltoni and T.~Schwetz, \emph{{Global Analyses of
  Neutrino Oscillation Experiments}},
  \href{http://arxiv.org/abs/1512.06856}{{\tt 1512.06856}}.

\bibitem{Ade:2013zuv}
{\scshape Planck} collaboration, P.~A.~R. Ade et~al., \emph{{Planck 2013
  results. XVI. Cosmological parameters}},
  \href{http://dx.doi.org/10.1051/0004-6361/201321591}{\emph{Astron.
  Astrophys.} {\bf 571} (2014) A16},
  [\href{http://arxiv.org/abs/1303.5076}{{\tt 1303.5076}}].

\bibitem{Minkowski:1977sc}
P.~Minkowski, \emph{{mu $\rightarrow$ e gamma at a Rate of One Out of 1-Billion
  Muon Decays?}},
  \href{http://dx.doi.org/10.1016/0370-2693(77)90435-X}{\emph{Phys.Lett.} {\bf
  B67} (1977) 421}.

\bibitem{Yanagida:see-saw}
T.~Yanagida, \emph{{HORIZONTAL SYMMETRY AND MASSES OF NEUTRINOS}},
  {\emph{Conf.Proc.} {\bf C7902131} (1979) 95--99}.

\bibitem{Gell-Mann:see-saw}
M.~Gell-Mann, P.~Ramond and R.~Slansky, \emph{{Complex Spinors and Unified
  Theories}}, {\emph{Conf.Proc.} {\bf C790927} (1979) 315--321},
  [\href{http://arxiv.org/abs/1306.4669}{{\tt 1306.4669}}].

\bibitem{Mohapatra:1979ia}
R.~N. Mohapatra and G.~Senjanovic, \emph{{Neutrino Mass and Spontaneous Parity
  Violation}},
  \href{http://dx.doi.org/10.1103/PhysRevLett.44.912}{\emph{Phys.Rev.Lett.}
  {\bf 44} (1980) 912}.

\bibitem{Cheng:1980qt}
T.~P. Cheng and L.-F. Li, \emph{{Neutrino Masses, Mixings and Oscillations in
  SU(2) x U(1) Models of Electroweak Interactions}},
  \href{http://dx.doi.org/10.1103/PhysRevD.22.2860}{\emph{Phys. Rev.} {\bf D22}
  (1980) 2860}.

\bibitem{Schechter:1980gr}
J.~Schechter and J.~Valle, \emph{{Neutrino Masses in SU(2) x U(1) Theories}},
  \href{http://dx.doi.org/10.1103/PhysRevD.22.2227}{\emph{Phys.Rev.} {\bf D22}
  (1980) 2227}.

\bibitem{Schechter:1981cv}
J.~Schechter and J.~Valle, \emph{{Neutrino Decay and Spontaneous Violation of
  Lepton Number}},
  \href{http://dx.doi.org/10.1103/PhysRevD.25.774}{\emph{Phys.Rev.} {\bf D25}
  (1982) 774}.

\bibitem{Zee:1980ai}
A.~Zee, \emph{{A Theory of Lepton Number Violation, Neutrino Majorana Mass, and
  Oscillation}},
  \href{http://dx.doi.org/10.1016/0370-2693(80)90349-4}{\emph{Phys. Lett.} {\bf
  B93} (1980) 389}.

\bibitem{Mohapatra:1998rq}
R.~N. Mohapatra and P.~B. Pal, \emph{{Massive neutrinos in physics and
  astrophysics. Second edition}}, {\emph{World Sci. Lect. Notes Phys.} {\bf 60}
  (1998) 1--397}.

\bibitem{Fukugita:2003en}
M.~Fukugita and T.~Yanagida, \emph{{Physics of neutrinos and applications to
  astrophysics}}.
\newblock 2003.

\bibitem{Ma:2009wt}
E.~Ma, \emph{{Neutrino Theory: Some Recent Developments}},  in \emph{{Particles
  and fields. Proceedings, Meeting of the Division of the American Physical
  Society, DPF 2009, Detroit, USA, July 26-31, 2009}}, 2009.
\newblock \href{http://arxiv.org/abs/0908.1770}{{\tt 0908.1770}}.

\bibitem{Altarelli:2014dca}
G.~Altarelli, \emph{{Status of Neutrino Mass and Mixing}},
  \href{http://dx.doi.org/10.1142/S0217751X14440023}{\emph{Int. J. Mod. Phys.}
  {\bf A29} (2014) 1444002}, [\href{http://arxiv.org/abs/1404.3859}{{\tt
  1404.3859}}].

\bibitem{Zee:1985id}
A.~Zee, \emph{{Quantum Numbers of Majorana Neutrino Masses}},
  \href{http://dx.doi.org/10.1016/0550-3213(86)90475-X}{\emph{Nucl. Phys.} {\bf
  B264} (1986) 99}.

\bibitem{Babu:1988ki}
K.~S. Babu, \emph{{Model of 'Calculable' Majorana Neutrino Masses}},
  \href{http://dx.doi.org/10.1016/0370-2693(88)91584-5}{\emph{Phys. Lett.} {\bf
  B203} (1988) 132}.

\bibitem{Krauss:2002px}
L.~M. Krauss, S.~Nasri and M.~Trodden, \emph{{A Model for neutrino masses and
  dark matter}},
  \href{http://dx.doi.org/10.1103/PhysRevD.67.085002}{\emph{Phys. Rev.} {\bf
  D67} (2003) 085002}, [\href{http://arxiv.org/abs/hep-ph/0210389}{{\tt
  hep-ph/0210389}}].

\bibitem{Ma:2006km}
E.~Ma, \emph{{Verifiable radiative seesaw mechanism of neutrino mass and dark
  matter}}, \href{http://dx.doi.org/10.1103/PhysRevD.73.077301}{\emph{Phys.
  Rev.} {\bf D73} (2006) 077301},
  [\href{http://arxiv.org/abs/hep-ph/0601225}{{\tt hep-ph/0601225}}].

\bibitem{Aoki:2008av}
M.~Aoki, S.~Kanemura and O.~Seto, \emph{{Neutrino mass, Dark Matter and Baryon
  Asymmetry via TeV-Scale Physics without Fine-Tuning}},
  \href{http://dx.doi.org/10.1103/PhysRevLett.102.051805}{\emph{Phys. Rev.
  Lett.} {\bf 102} (2009) 051805}, [\href{http://arxiv.org/abs/0807.0361}{{\tt
  0807.0361}}].

\bibitem{Gustafsson:2012vj}
M.~Gustafsson, J.~M. No and M.~A. Rivera, \emph{{Predictive Model for
  Radiatively Induced Neutrino Masses and Mixings with Dark Matter}},
  \href{http://dx.doi.org/10.1103/PhysRevLett.110.211802,
  10.1103/PhysRevLett.112.259902}{\emph{Phys. Rev. Lett.} {\bf 110} (2013)
  211802}, [\href{http://arxiv.org/abs/1212.4806}{{\tt 1212.4806}}].

\bibitem{Hatanaka:2014tba}
H.~Hatanaka, K.~Nishiwaki, H.~Okada and Y.~Orikasa, \emph{{A Three-Loop
  Neutrino Model with Global $U(1)$ Symmetry}},
  \href{http://dx.doi.org/10.1016/j.nuclphysb.2015.03.006}{\emph{Nucl. Phys.}
  {\bf B894} (2015) 268--283}, [\href{http://arxiv.org/abs/1412.8664}{{\tt
  1412.8664}}].

\bibitem{Nishiwaki:2015iqa}
K.~Nishiwaki, H.~Okada and Y.~Orikasa, \emph{{Three loop neutrino model with
  isolated $k^{\pm\pm}$}},
  \href{http://dx.doi.org/10.1103/PhysRevD.92.093013}{\emph{Phys. Rev.} {\bf
  D92} (2015) 093013}, [\href{http://arxiv.org/abs/1507.02412}{{\tt
  1507.02412}}].

\bibitem{Chao:2015nac}
W.~Chao, \emph{{Neutrino Catalyzed Diphoton Excess}},
  \href{http://arxiv.org/abs/1512.08484}{{\tt 1512.08484}}.

\bibitem{Kanemura:2015bli}
S.~Kanemura, K.~Nishiwaki, H.~Okada, Y.~Orikasa, S.~C. Park and R.~Watanabe,
  \emph{{LHC 750 GeV Diphoton excess in a radiative seesaw model}},
  \href{http://arxiv.org/abs/1512.09048}{{\tt 1512.09048}}.

\bibitem{Nomura:2016fzs}
T.~Nomura and H.~Okada, \emph{{Four-loop Neutrino Model Inspired by Diphoton
  Excess at 750 GeV}},  \href{http://arxiv.org/abs/1601.00386}{{\tt
  1601.00386}}.

\bibitem{Yu:2016lof}
J.-H. Yu, \emph{{Hidden Gauged U(1) Model: Unifying Scotogenic Neutrino and
  Flavor Dark Matter}},  \href{http://arxiv.org/abs/1601.02609}{{\tt
  1601.02609}}.

\bibitem{Ding:2016ldt}
R.~Ding, Z.-L. Han, Y.~Liao and X.-D. Ma, \emph{{Interpretation of 750 GeV
  Diphoton Excess at LHC in Singlet Extension of Color-octet Neutrino Mass
  Model}},  \href{http://arxiv.org/abs/1601.02714}{{\tt 1601.02714}}.

\bibitem{Nomura:2016seu}
T.~Nomura and H.~Okada, \emph{{Four-loop Radiative Seesaw Model with 750 GeV
  Diphoton Resonance}},  \href{http://arxiv.org/abs/1601.04516}{{\tt
  1601.04516}}.

\bibitem{Okada:2016rav}
H.~Okada and K.~Yagyu, \emph{{Renormalizable Model for Neutrino Mass, Dark
  Matter, Muon $g-2$ and 750 GeV Diphoton Excess}},
  \href{http://arxiv.org/abs/1601.05038}{{\tt 1601.05038}}.

\bibitem{KlapdorKleingrothaus:2000sn}
H.~V. Klapdor-Kleingrothaus et~al., \emph{{Latest results from the
  Heidelberg-Moscow double beta decay experiment}},
  \href{http://dx.doi.org/10.1007/s100500170022}{\emph{Eur. Phys. J.} {\bf A12}
  (2001) 147--154}, [\href{http://arxiv.org/abs/hep-ph/0103062}{{\tt
  hep-ph/0103062}}].

\bibitem{KlapdorKleingrothaus:2001ke}
H.~V. Klapdor-Kleingrothaus, A.~Dietz, H.~L. Harney and I.~V. Krivosheina,
  \emph{{Evidence for neutrinoless double beta decay}},
  \href{http://dx.doi.org/10.1142/S0217732301005825}{\emph{Mod. Phys. Lett.}
  {\bf A16} (2001) 2409--2420},
  [\href{http://arxiv.org/abs/hep-ph/0201231}{{\tt hep-ph/0201231}}].

\bibitem{Agashe:2014kda}
{\scshape Particle Data Group} collaboration, K.~A. Olive et~al., \emph{{Review
  of Particle Physics}},
  \href{http://dx.doi.org/10.1088/1674-1137/38/9/090001}{\emph{Chin. Phys.}
  {\bf C38} (2014) 090001}.

\bibitem{ArkaniHamed:1998rs}
N.~Arkani-Hamed, S.~Dimopoulos and G.~R. Dvali, \emph{{The Hierarchy problem
  and new dimensions at a millimeter}},
  \href{http://dx.doi.org/10.1016/S0370-2693(98)00466-3}{\emph{Phys. Lett.}
  {\bf B429} (1998) 263--272}, [\href{http://arxiv.org/abs/hep-ph/9803315}{{\tt
  hep-ph/9803315}}].

\bibitem{Antoniadis:1998ig}
I.~Antoniadis, N.~Arkani-Hamed, S.~Dimopoulos and G.~R. Dvali, \emph{{New
  dimensions at a millimeter to a Fermi and superstrings at a TeV}},
  \href{http://dx.doi.org/10.1016/S0370-2693(98)00860-0}{\emph{Phys. Lett.}
  {\bf B436} (1998) 257--263}, [\href{http://arxiv.org/abs/hep-ph/9804398}{{\tt
  hep-ph/9804398}}].

\bibitem{Gogberashvili:1998vx}
M.~Gogberashvili, \emph{{Hierarchy problem in the shell universe model}},
  \href{http://dx.doi.org/10.1142/S0218271802002992}{\emph{Int. J. Mod. Phys.}
  {\bf D11} (2002) 1635--1638},
  [\href{http://arxiv.org/abs/hep-ph/9812296}{{\tt hep-ph/9812296}}].

\bibitem{Randall:1999ee}
L.~Randall and R.~Sundrum, \emph{{A Large mass hierarchy from a small extra
  dimension}}, \href{http://dx.doi.org/10.1103/PhysRevLett.83.3370}{\emph{Phys.
  Rev. Lett.} {\bf 83} (1999) 3370--3373},
  [\href{http://arxiv.org/abs/hep-ph/9905221}{{\tt hep-ph/9905221}}].

\bibitem{ArkaniHamed:1998vp}
N.~Arkani-Hamed, S.~Dimopoulos, G.~R. Dvali and J.~March-Russell,
  \emph{{Neutrino masses from large extra dimensions}},
  \href{http://dx.doi.org/10.1103/PhysRevD.65.024032}{\emph{Phys.Rev.} {\bf
  D65} (2002) 024032}, [\href{http://arxiv.org/abs/hep-ph/9811448}{{\tt
  hep-ph/9811448}}].

\bibitem{ArkaniHamed:1999dc}
N.~Arkani-Hamed and M.~Schmaltz, \emph{{Hierarchies without symmetries from
  extra dimensions}},
  \href{http://dx.doi.org/10.1103/PhysRevD.61.033005}{\emph{Phys.Rev.} {\bf
  D61} (2000) 033005}, [\href{http://arxiv.org/abs/hep-ph/9903417}{{\tt
  hep-ph/9903417}}].

\bibitem{Dvali:1999cn}
G.~Dvali and A.~Y. Smirnov, \emph{{Probing large extra dimensions with
  neutrinos}},
  \href{http://dx.doi.org/10.1016/S0550-3213(99)00574-X}{\emph{Nucl.Phys.} {\bf
  B563} (1999) 63--81}, [\href{http://arxiv.org/abs/hep-ph/9904211}{{\tt
  hep-ph/9904211}}].

\bibitem{Yoshioka:1999ds}
K.~Yoshioka, \emph{{On fermion mass hierarchy with extra dimensions}},
  \href{http://dx.doi.org/10.1016/S0217-7323(00)00006-2}{\emph{Mod.Phys.Lett.}
  {\bf A15} (2000) 29--40}, [\href{http://arxiv.org/abs/hep-ph/9904433}{{\tt
  hep-ph/9904433}}].

\bibitem{Mohapatra:1999zd}
R.~Mohapatra, S.~Nandi and A.~Perez-Lorenzana, \emph{{Neutrino masses and
  oscillations in models with large extra dimensions}},
  \href{http://dx.doi.org/10.1016/S0370-2693(99)01119-3}{\emph{Phys.Lett.} {\bf
  B466} (1999) 115--121}, [\href{http://arxiv.org/abs/hep-ph/9907520}{{\tt
  hep-ph/9907520}}].

\bibitem{Grossman:1999ra}
Y.~Grossman and M.~Neubert, \emph{{Neutrino masses and mixings in
  nonfactorizable geometry}},
  \href{http://dx.doi.org/10.1016/S0370-2693(00)00054-X}{\emph{Phys.Lett.} {\bf
  B474} (2000) 361--371}, [\href{http://arxiv.org/abs/hep-ph/9912408}{{\tt
  hep-ph/9912408}}].

\bibitem{Gherghetta:2000qt}
T.~Gherghetta and A.~Pomarol, \emph{{Bulk fields and supersymmetry in a slice
  of AdS}},
  \href{http://dx.doi.org/10.1016/S0550-3213(00)00392-8}{\emph{Nucl.Phys.} {\bf
  B586} (2000) 141--162}, [\href{http://arxiv.org/abs/hep-ph/0003129}{{\tt
  hep-ph/0003129}}].

\bibitem{Huber:2000ie}
S.~J. Huber and Q.~Shafi, \emph{{Fermion masses, mixings and proton decay in a
  Randall-Sundrum model}},
  \href{http://dx.doi.org/10.1016/S0370-2693(00)01399-X}{\emph{Phys.Lett.} {\bf
  B498} (2001) 256--262}, [\href{http://arxiv.org/abs/hep-ph/0010195}{{\tt
  hep-ph/0010195}}].

\bibitem{Moreau:2004qe}
G.~Moreau, \emph{{Realistic neutrino masses from multi-brane extensions of the
  Randall-Sundrum model?}},
  \href{http://dx.doi.org/10.1140/epjc/s2005-02165-5}{\emph{Eur. Phys. J.} {\bf
  C40} (2005) 539--554}, [\href{http://arxiv.org/abs/hep-ph/0407177}{{\tt
  hep-ph/0407177}}].

\bibitem{Moreau:2005kz}
G.~Moreau and J.~I. Silva-Marcos, \emph{{Neutrinos in warped extra
  dimensions}},
  \href{http://dx.doi.org/10.1088/1126-6708/2006/01/048}{\emph{JHEP} {\bf 01}
  (2006) 048}, [\href{http://arxiv.org/abs/hep-ph/0507145}{{\tt
  hep-ph/0507145}}].

\bibitem{Frank:2014aca}
M.~Frank, C.~Hamzaoui, N.~Pourtolami and M.~Toharia, \emph{{Unified Flavor
  Symmetry from warped dimensions}},
  \href{http://dx.doi.org/10.1016/j.physletb.2015.01.025}{\emph{Phys. Lett.}
  {\bf B742} (2015) 178--182}, [\href{http://arxiv.org/abs/1406.2331}{{\tt
  1406.2331}}].

\bibitem{Frank:2015sua}
M.~Frank, C.~Hamzaoui, N.~Pourtolami and M.~Toharia, \emph{{Fermion Masses and
  Mixing in General Warped Extra Dimensional Models}},
  \href{http://dx.doi.org/10.1103/PhysRevD.91.116001}{\emph{Phys. Rev.} {\bf
  D91} (2015) 116001}, [\href{http://arxiv.org/abs/1504.02780}{{\tt
  1504.02780}}].

\bibitem{Dvali:2000ha}
G.~R. Dvali and M.~A. Shifman, \emph{{Families as neighbors in extra
  dimension}},
  \href{http://dx.doi.org/10.1016/S0370-2693(00)00083-6}{\emph{Phys.Lett.} {\bf
  B475} (2000) 295--302}, [\href{http://arxiv.org/abs/hep-ph/0001072}{{\tt
  hep-ph/0001072}}].

\bibitem{Shaposhnikov:2001nz}
M.~E. Shaposhnikov and P.~Tinyakov, \emph{{Extra dimensions as an alternative
  to Higgs mechanism?}},
  \href{http://dx.doi.org/10.1016/S0370-2693(01)00781-X}{\emph{Phys. Lett.}
  {\bf B515} (2001) 442--446}, [\href{http://arxiv.org/abs/hep-th/0102161}{{\tt
  hep-th/0102161}}].

\bibitem{Neronov:2001qv}
A.~Neronov, \emph{{Fermion masses and quantum numbers from extra dimensions}},
  \href{http://dx.doi.org/10.1103/PhysRevD.65.044004}{\emph{Phys.Rev.} {\bf
  D65} (2002) 044004}, [\href{http://arxiv.org/abs/gr-qc/0106092}{{\tt
  gr-qc/0106092}}].

\bibitem{Kaplan:2001ga}
D.~E. Kaplan and T.~M.~P. Tait, \emph{{New tools for fermion masses from extra
  dimensions}}, {\emph{JHEP} {\bf 0111} (2001) 051},
  [\href{http://arxiv.org/abs/hep-ph/0110126}{{\tt hep-ph/0110126}}].

\bibitem{Libanov:2000uf}
M.~V. Libanov and S.~V. Troitsky, \emph{{Three fermionic generations on a
  topological defect in extra dimensions}},
  \href{http://dx.doi.org/10.1016/S0550-3213(01)00036-0}{\emph{Nucl.Phys.} {\bf
  B599} (2001) 319--333}, [\href{http://arxiv.org/abs/hep-ph/0011095}{{\tt
  hep-ph/0011095}}].

\bibitem{Frere:2000dc}
J.~M. Frere, M.~V. Libanov and S.~V. Troitsky, \emph{{Three generations on a
  local vortex in extra dimensions}},
  \href{http://dx.doi.org/10.1016/S0370-2693(01)00696-7}{\emph{Phys.Lett.} {\bf
  B512} (2001) 169--173}, [\href{http://arxiv.org/abs/hep-ph/0012306}{{\tt
  hep-ph/0012306}}].

\bibitem{Frere:2001ug}
J.~M. Frere, M.~V. Libanov and S.~V. Troitsky, \emph{{Neutrino masses with a
  single generation in the bulk}}, {\emph{JHEP} {\bf 0111} (2001) 025},
  [\href{http://arxiv.org/abs/hep-ph/0110045}{{\tt hep-ph/0110045}}].

\bibitem{Frere:2003hn}
J.~M. Frere, G.~Moreau and E.~Nezri, \emph{{Neutrino mass patterns within the
  seesaw model from multilocalization along extra dimensions}},
  \href{http://dx.doi.org/10.1103/PhysRevD.69.033003}{\emph{Phys. Rev.} {\bf
  D69} (2004) 033003}, [\href{http://arxiv.org/abs/hep-ph/0309218}{{\tt
  hep-ph/0309218}}].

\bibitem{Cremades:2004wa}
D.~Cremades, L.~Ibanez and F.~Marchesano, \emph{{Computing Yukawa couplings
  from magnetized extra dimensions}},
  \href{http://dx.doi.org/10.1088/1126-6708/2004/05/079}{\emph{JHEP} {\bf 0405}
  (2004) 079}, [\href{http://arxiv.org/abs/hep-th/0404229}{{\tt
  hep-th/0404229}}].

\bibitem{Nagasawa:2004xk}
T.~Nagasawa and M.~Sakamoto, \emph{{Higgsless gauge symmetry breaking with a
  large mass hierarchy}},
  \href{http://dx.doi.org/10.1143/PTP.112.629}{\emph{Prog. Theor. Phys.} {\bf
  112} (2004) 629--638}, [\href{http://arxiv.org/abs/hep-ph/0406024}{{\tt
  hep-ph/0406024}}].

\bibitem{Parameswaran:2006db}
S.~L. Parameswaran, S.~Randjbar-Daemi and A.~Salvio, \emph{{Gauge Fields,
  Fermions and Mass Gaps in 6D Brane Worlds}},
  \href{http://dx.doi.org/10.1016/j.nuclphysb.2006.12.020}{\emph{Nucl.Phys.}
  {\bf B767} (2007) 54--81}, [\href{http://arxiv.org/abs/hep-th/0608074}{{\tt
  hep-th/0608074}}].

\bibitem{Abada:2006yd}
A.~Abada, P.~Dey and G.~Moreau, \emph{{Neutrinos in flat extra dimension:
  Towards a realistic scenario}},
  \href{http://dx.doi.org/10.1088/1126-6708/2007/09/006}{\emph{JHEP} {\bf 09}
  (2007) 006}, [\href{http://arxiv.org/abs/hep-ph/0611200}{{\tt
  hep-ph/0611200}}].

\bibitem{Gogberashvili:2007gg}
M.~Gogberashvili, P.~Midodashvili and D.~Singleton, \emph{{Fermion Generations
  from 'Apple-Shaped' Extra Dimensions}},
  \href{http://dx.doi.org/10.1088/1126-6708/2007/08/033}{\emph{JHEP} {\bf 0708}
  (2007) 033}, [\href{http://arxiv.org/abs/0706.0676}{{\tt 0706.0676}}].

\bibitem{Park:2009cs}
S.~C. Park and J.~Shu, \emph{{Split Universal Extra Dimensions and Dark
  Matter}}, \href{http://dx.doi.org/10.1103/PhysRevD.79.091702}{\emph{Phys.
  Rev.} {\bf D79} (2009) 091702}, [\href{http://arxiv.org/abs/0901.0720}{{\tt
  0901.0720}}].

\bibitem{Kusenko:2010ik}
A.~Kusenko, F.~Takahashi and T.~T. Yanagida, \emph{{Dark Matter from Split
  Seesaw}}, \href{http://dx.doi.org/10.1016/j.physletb.2010.08.031}{\emph{Phys.
  Lett.} {\bf B693} (2010) 144--148},
  [\href{http://arxiv.org/abs/1006.1731}{{\tt 1006.1731}}].

\bibitem{Csaki:2010az}
C.~Csaki, J.~Heinonen, J.~Hubisz, S.~C. Park and J.~Shu, \emph{{5D UED: Flat
  and Flavorless}},
  \href{http://dx.doi.org/10.1007/JHEP01(2011)089}{\emph{JHEP} {\bf 01} (2011)
  089}, [\href{http://arxiv.org/abs/1007.0025}{{\tt 1007.0025}}].

\bibitem{Fujimoto:2011kf}
Y.~Fujimoto, T.~Nagasawa, S.~Ohya and M.~Sakamoto, \emph{{Phase Structure of
  Gauge Theories on an Interval}},
  \href{http://dx.doi.org/10.1143/PTP.126.841}{\emph{Prog. Theor. Phys.} {\bf
  126} (2011) 841--854}, [\href{http://arxiv.org/abs/1108.1976}{{\tt
  1108.1976}}].

\bibitem{Kaplan:2011vz}
D.~B. Kaplan and S.~Sun, \emph{{Spacetime as a topological insulator: Mechanism
  for the origin of the fermion generations}},
  \href{http://dx.doi.org/10.1103/PhysRevLett.108.181807,
  10.1103/PhysRevLett.108.209901}{\emph{Phys.Rev.Lett.} {\bf 108} (2012)
  181807}, [\href{http://arxiv.org/abs/1112.0302}{{\tt 1112.0302}}].

\bibitem{Fujimoto:2012wv}
Y.~Fujimoto, T.~Nagasawa, K.~Nishiwaki and M.~Sakamoto, \emph{{Quark mass
  hierarchy and mixing via geometry of extra dimension with point
  interactions}}, \href{http://dx.doi.org/10.1093/ptep/pts097}{\emph{PTEP} {\bf
  2013} (2013) 023B07}, [\href{http://arxiv.org/abs/1209.5150}{{\tt
  1209.5150}}].

\bibitem{Fujimoto:2013ki}
Y.~Fujimoto, K.~Nishiwaki and M.~Sakamoto, \emph{{CP phase from twisted Higgs
  VEV in extra dimension}},
  \href{http://dx.doi.org/10.1103/PhysRevD.88.115007}{\emph{Phys.Rev.} {\bf
  D88} (2013) 115007}, [\href{http://arxiv.org/abs/1301.7253}{{\tt
  1301.7253}}].

\bibitem{Fujimoto:2013xha}
Y.~Fujimoto, T.~Kobayashi, T.~Miura, K.~Nishiwaki and M.~Sakamoto,
  \emph{{Shifted orbifold models with magnetic flux}},
  \href{http://dx.doi.org/10.1103/PhysRevD.87.086001}{\emph{Phys.Rev.} {\bf
  D87} (2013) 086001}, [\href{http://arxiv.org/abs/1302.5768}{{\tt
  1302.5768}}].

\bibitem{Takahashi:2013eva}
R.~Takahashi, \emph{{Separate seesaw and its applications to dark matter and
  baryogenesis}}, \href{http://dx.doi.org/10.1093/ptep/ptt042}{\emph{PTEP} {\bf
  2013} (2013) 063B04}, [\href{http://arxiv.org/abs/1303.0108}{{\tt
  1303.0108}}].

\bibitem{Abe:2013bca}
T.-H. Abe, Y.~Fujimoto, T.~Kobayashi, T.~Miura, K.~Nishiwaki et~al.,
  \emph{{$Z_N$ twisted orbifold models with magnetic flux}},
  \href{http://dx.doi.org/10.1007/JHEP01(2014)065}{\emph{JHEP} {\bf 1401}
  (2014) 065}, [\href{http://arxiv.org/abs/1309.4925}{{\tt 1309.4925}}].

\bibitem{Cai:2015jla}
C.~Cai and H.-H. Zhang, \emph{{Majorana neutrinos with point interactions}},
  \href{http://arxiv.org/abs/1503.08805}{{\tt 1503.08805}}.

\bibitem{Fujimoto:2014fka}
Y.~Fujimoto, K.~Nishiwaki, M.~Sakamoto and R.~Takahashi, \emph{{Realization of
  lepton masses and mixing angles from point interactions in an extra
  dimension}}, \href{http://dx.doi.org/10.1007/JHEP10(2014)191}{\emph{JHEP}
  {\bf 10} (2014) 191}, [\href{http://arxiv.org/abs/1405.5872}{{\tt
  1405.5872}}].

\bibitem{Abe:2014noa}
T.-h. Abe, Y.~Fujimoto, T.~Kobayashi, T.~Miura, K.~Nishiwaki and M.~Sakamoto,
  \emph{{Operator analysis of physical states on magnetized $T^{2}/Z_{N}$
  orbifolds}},
  \href{http://dx.doi.org/10.1016/j.nuclphysb.2014.11.022}{\emph{Nucl. Phys.}
  {\bf B890} (2014) 442--480}, [\href{http://arxiv.org/abs/1409.5421}{{\tt
  1409.5421}}].

\bibitem{Abe:2015yva}
T.-h. Abe, Y.~Fujimoto, T.~Kobayashi, T.~Miura, K.~Nishiwaki, M.~Sakamoto
  et~al., \emph{{Classification of three-generation models on magnetized
  orbifolds}},
  \href{http://dx.doi.org/10.1016/j.nuclphysb.2015.03.004}{\emph{Nucl. Phys.}
  {\bf B894} (2015) 374--406}, [\href{http://arxiv.org/abs/1501.02787}{{\tt
  1501.02787}}].

\bibitem{Barate:1999ce}
{\scshape ALEPH} collaboration, R.~Barate et~al., \emph{{Measurement of the Z
  resonance parameters at LEP}},
  \href{http://dx.doi.org/10.1007/s100520050732}{\emph{Eur. Phys. J.} {\bf C14}
  (2000) 1--50}.

\bibitem{Acciarri:2000ai}
{\scshape L3} collaboration, M.~Acciarri et~al., \emph{{Measurements of
  cross-sections and forward backward asymmetries at the $Z$ resonance and
  determination of electroweak parameters}},
  \href{http://dx.doi.org/10.1007/s100520050001}{\emph{Eur. Phys. J.} {\bf C16}
  (2000) 1--40}, [\href{http://arxiv.org/abs/hep-ex/0002046}{{\tt
  hep-ex/0002046}}].

\bibitem{Abreu:2000mh}
{\scshape DELPHI} collaboration, P.~Abreu et~al., \emph{{Cross-sections and
  leptonic forward backward asymmetries from the Z0 running of LEP}},
  \href{http://dx.doi.org/10.1007/s100520000392}{\emph{Eur. Phys. J.} {\bf C16}
  (2000) 371--405}.

\bibitem{Abbiendi:2000hu}
{\scshape OPAL} collaboration, G.~Abbiendi et~al., \emph{{Precise determination
  of the Z resonance parameters at LEP: 'Zedometry'}},
  \href{http://dx.doi.org/10.1007/s100520100627}{\emph{Eur. Phys. J.} {\bf C19}
  (2001) 587--651}, [\href{http://arxiv.org/abs/hep-ex/0012018}{{\tt
  hep-ex/0012018}}].

\bibitem{Collaborations:2000aa}
{\scshape Line Shape Sub-Group of the LEP Electroweak Working Group, DELPHI,
  LEP, ALEPH, OPAL, L3} collaboration, \emph{{Combination procedure for the
  precise determination of Z boson parameters from results of the LEP
  experiments}},  \href{http://arxiv.org/abs/hep-ex/0101027}{{\tt
  hep-ex/0101027}}.

\bibitem{ALEPH:2002aa}
{\scshape SLD Heavy Flavor Group, DELPHI, ALEPH, OPAL, LEP Electroweak Working
  Group, L3} collaboration, \emph{{A Combination of preliminary electroweak
  measurements and constraints on the standard model}},
  \href{http://arxiv.org/abs/hep-ex/0212036}{{\tt hep-ex/0212036}}.

\bibitem{Lim:2005rc}
C.~S. Lim, T.~Nagasawa, M.~Sakamoto and H.~Sonoda, \emph{{Supersymmetry in
  gauge theories with extra dimensions}},
  \href{http://dx.doi.org/10.1103/PhysRevD.72.064006}{\emph{Phys.Rev.} {\bf
  D72} (2005) 064006}, [\href{http://arxiv.org/abs/hep-th/0502022}{{\tt
  hep-th/0502022}}].

\bibitem{Lim:2007fy}
C.~S. Lim, T.~Nagasawa, S.~Ohya, K.~Sakamoto and M.~Sakamoto,
  \emph{{Supersymmetry in 5d gravity}},
  \href{http://dx.doi.org/10.1103/PhysRevD.77.045020}{\emph{Phys.Rev.} {\bf
  D77} (2008) 045020}, [\href{http://arxiv.org/abs/0710.0170}{{\tt
  0710.0170}}].

\bibitem{Lim:2008hi}
C.~S. Lim, T.~Nagasawa, S.~Ohya, K.~Sakamoto and M.~Sakamoto,
  \emph{{Gauge-Fixing and Residual Symmetries in Gauge/Gravity Theories with
  Extra Dimensions}},
  \href{http://dx.doi.org/10.1103/PhysRevD.77.065009}{\emph{Phys.Rev.} {\bf
  D77} (2008) 065009}, [\href{http://arxiv.org/abs/0801.0845}{{\tt
  0801.0845}}].

\bibitem{Nagasawa:2008an}
T.~Nagasawa, S.~Ohya, K.~Sakamoto, M.~Sakamoto and K.~Sekiya, \emph{{Hierarchy
  of QM SUSYs on a Bounded Domain}},
  \href{http://dx.doi.org/10.1088/1751-8113/42/26/265203}{\emph{J.Phys.} {\bf
  A42} (2009) 265203}, [\href{http://arxiv.org/abs/0812.4659}{{\tt
  0812.4659}}].

\bibitem{Csaki:2003dt}
C.~Csaki, C.~Grojean, H.~Murayama, L.~Pilo and J.~Terning, \emph{{Gauge
  theories on an interval: Unitarity without a Higgs boson}},
  \href{http://dx.doi.org/10.1103/PhysRevD.69.055006}{\emph{Phys.Rev.} {\bf
  D69} (2004) 055006}, [\href{http://arxiv.org/abs/hep-ph/0305237}{{\tt
  hep-ph/0305237}}].

\bibitem{Csaki:2003sh}
C.~Csaki, C.~Grojean, J.~Hubisz, Y.~Shirman and J.~Terning, \emph{{Fermions on
  an interval: Quark and lepton masses without a Higgs}},
  \href{http://dx.doi.org/10.1103/PhysRevD.70.015012}{\emph{Phys. Rev.} {\bf
  D70} (2004) 015012}, [\href{http://arxiv.org/abs/hep-ph/0310355}{{\tt
  hep-ph/0310355}}].

\bibitem{Csaki:2005vy}
C.~Csaki, J.~Hubisz and P.~Meade, \emph{{TASI lectures on electroweak symmetry
  breaking from extra dimensions}},  in \emph{{Physics in $D >= 4$.
  Proceedings, Theoretical Advanced Study Institute in elementary particle
  physics, TASI 2004, Boulder, USA, June 6-July 2, 2004}}, pp.~703--776, 2005.
\newblock \href{http://arxiv.org/abs/hep-ph/0510275}{{\tt hep-ph/0510275}}.

\bibitem{Donini:2015ejd}
A.~Donini, \emph{{A scalar field coupled to a brane in ${\cal M}_4 \times {\cal
  S}_1$. Part I: Kaluza-Klein spectrum and zero-mode localization}},
  \href{http://arxiv.org/abs/1512.03978}{{\tt 1512.03978}}.

\bibitem{Peskin:1990zt}
M.~E. Peskin and T.~Takeuchi, \emph{{A New constraint on a strongly interacting
  Higgs sector}},
  \href{http://dx.doi.org/10.1103/PhysRevLett.65.964}{\emph{Phys. Rev. Lett.}
  {\bf 65} (1990) 964--967}.

\bibitem{Peskin:1991sw}
M.~E. Peskin and T.~Takeuchi, \emph{{Estimation of oblique electroweak
  corrections}}, \href{http://dx.doi.org/10.1103/PhysRevD.46.381}{\emph{Phys.
  Rev.} {\bf D46} (1992) 381--409}.

\bibitem{Rizzo:1999br}
T.~G. Rizzo and J.~D. Wells, \emph{{Electroweak precision measurements and
  collider probes of the standard model with large extra dimensions}},
  \href{http://dx.doi.org/10.1103/PhysRevD.61.016007}{\emph{Phys. Rev.} {\bf
  D61} (2000) 016007}, [\href{http://arxiv.org/abs/hep-ph/9906234}{{\tt
  hep-ph/9906234}}].

\bibitem{Davoudiasl:1999tf}
H.~Davoudiasl, J.~L. Hewett and T.~G. Rizzo, \emph{{Bulk gauge fields in the
  Randall-Sundrum model}},
  \href{http://dx.doi.org/10.1016/S0370-2693(99)01430-6}{\emph{Phys. Lett.}
  {\bf B473} (2000) 43--49}, [\href{http://arxiv.org/abs/hep-ph/9911262}{{\tt
  hep-ph/9911262}}].

\bibitem{Csaki:2002gy}
C.~Csaki, J.~Erlich and J.~Terning, \emph{{The Effective Lagrangian in the
  Randall-Sundrum model and electroweak physics}},
  \href{http://dx.doi.org/10.1103/PhysRevD.66.064021}{\emph{Phys. Rev.} {\bf
  D66} (2002) 064021}, [\href{http://arxiv.org/abs/hep-ph/0203034}{{\tt
  hep-ph/0203034}}].

\bibitem{Flacke:2011nb}
T.~Flacke and C.~Pasold, \emph{{Constraints on split-UED from Electroweak
  Precision Tests}},
  \href{http://dx.doi.org/10.1103/PhysRevD.85.126007}{\emph{Phys. Rev.} {\bf
  D85} (2012) 126007}, [\href{http://arxiv.org/abs/1111.7250}{{\tt
  1111.7250}}].

\bibitem{Baak:2014ora}
{\scshape Gfitter Group} collaboration, M.~Baak, J.~Cuth, J.~Haller,
  A.~Hoecker, R.~Kogler, K.~Monig et~al., \emph{{The global electroweak fit at
  NNLO and prospects for the LHC and ILC}},
  \href{http://dx.doi.org/10.1140/epjc/s10052-014-3046-5}{\emph{Eur. Phys. J.}
  {\bf C74} (2014) 3046}, [\href{http://arxiv.org/abs/1407.3792}{{\tt
  1407.3792}}].

\bibitem{Denner:1991kt}
A.~Denner, \emph{{Techniques for calculation of electroweak radiative
  corrections at the one loop level and results for W physics at LEP-200}},
  \href{http://dx.doi.org/10.1002/prop.2190410402}{\emph{Fortsch. Phys.} {\bf
  41} (1993) 307--420}, [\href{http://arxiv.org/abs/0709.1075}{{\tt
  0709.1075}}].

\bibitem{SekharChivukula:2001hz}
R.~S. Chivukula, D.~A. Dicus and H.-J. He, \emph{{Unitarity of compactified
  five-dimensional Yang-Mills theory}},
  \href{http://dx.doi.org/10.1016/S0370-2693(01)01435-6}{\emph{Phys.Lett.} {\bf
  B525} (2002) 175--182}, [\href{http://arxiv.org/abs/hep-ph/0111016}{{\tt
  hep-ph/0111016}}].

\bibitem{Abe:2003vg}
Y.~Abe, N.~Haba, Y.~Higashide, K.~Kobayashi and M.~Matsunaga, \emph{{Unitarity
  in gauge symmetry breaking on orbifold}},
  \href{http://dx.doi.org/10.1143/PTP.109.831}{\emph{Prog.Theor.Phys.} {\bf
  109} (2003) 831--842}, [\href{http://arxiv.org/abs/hep-th/0302115}{{\tt
  hep-th/0302115}}].

\bibitem{Chivukula:2003kq}
R.~S. Chivukula, D.~A. Dicus, H.-J. He and S.~Nandi, \emph{{Unitarity of the
  higher dimensional standard model}},
  \href{http://dx.doi.org/10.1016/S0370-2693(03)00553-7}{\emph{Phys.Lett.} {\bf
  B562} (2003) 109--117}, [\href{http://arxiv.org/abs/hep-ph/0302263}{{\tt
  hep-ph/0302263}}].

\bibitem{Ohl:2003dp}
T.~Ohl and C.~Schwinn, \emph{{Unitarity, BRST symmetry and ward identities in
  orbifold gauge theories}},
  \href{http://dx.doi.org/10.1103/PhysRevD.70.045019}{\emph{Phys.Rev.} {\bf
  D70} (2004) 045019}, [\href{http://arxiv.org/abs/hep-ph/0312263}{{\tt
  hep-ph/0312263}}].

\bibitem{Abe:2004wv}
Y.~Abe, N.~Haba, K.~Hayakawa, Y.~Matsumoto, M.~Matsunaga et~al., \emph{{4-D
  equivalence theorem and gauge symmetry on orbifold}},
  \href{http://dx.doi.org/10.1143/PTP.113.199}{\emph{Prog.Theor.Phys.} {\bf
  113} (2005) 199--213}, [\href{http://arxiv.org/abs/hep-th/0402146}{{\tt
  hep-th/0402146}}].

\bibitem{Sakai:2006qi}
N.~Sakai and N.~Uekusa, \emph{{Selecting gauge theories on an interval by 5D
  gauge transformations}},
  \href{http://dx.doi.org/10.1143/PTP.118.315}{\emph{Prog.Theor.Phys.} {\bf
  118} (2007) 315--335}, [\href{http://arxiv.org/abs/hep-th/0604121}{{\tt
  hep-th/0604121}}].

\bibitem{Nishiwaki:2010te}
K.~Nishiwaki and K.-y. Oda, \emph{{Unitarity in Dirichlet Higgs Model}},
  \href{http://dx.doi.org/10.1140/epjc/s10052-011-1786-z}{\emph{Eur.Phys.J.}
  {\bf C71} (2011) 1786}, [\href{http://arxiv.org/abs/1011.0405}{{\tt
  1011.0405}}].

\bibitem{Dolgov:2002wy}
A.~D. Dolgov, \emph{{Neutrinos in cosmology}},
  \href{http://dx.doi.org/10.1016/S0370-1573(02)00139-4}{\emph{Phys. Rept.}
  {\bf 370} (2002) 333--535}, [\href{http://arxiv.org/abs/hep-ph/0202122}{{\tt
  hep-ph/0202122}}].

\end{thebibliography}\endgroup

\end{document}